\documentclass[]{aa}
\usepackage[varg]{txfonts}
\usepackage{natbib}
\usepackage[final]{hyperref}
\usepackage{xcolor}
\usepackage{multirow}
\usepackage[switch]{lineno}
\usepackage{booktabs}
\usepackage{siunitx}
\usepackage{upgreek}
\usepackage[normalem]{ulem} 

\begin{document}


\title{Do galaxy mergers prefer under-dense environments?}

\author{
	U.~Sureshkumar \inst{\ref{aff:wits}, \ref{aff:oauj}} 
	\and A.~Durkalec \inst{\ref{aff:ncbj}} 
	\and A.~Pollo \inst{\ref{aff:oauj},\ref{aff:ncbj}}
	\and W.~J.~Pearson \inst{\ref{aff:ncbj}}
        \and D.~J.~Farrow \inst{\ref{aff:daim}, \ref{aff:eamilne}}
        \and A.~Narayanan \inst{\ref{aff:iist}}
        \and J.~Loveday \inst{\ref{aff:sussex}}
        \and E.~N.~Taylor \inst{\ref{aff:cas_swisburne}}
        \and L.~E.~Suelves \inst{\ref{aff:ncbj}}
	}

\institute{
    Wits Centre for Astrophysics, School of Physics, University of the Witwatersrand, Private Bag 3, Johannesburg 2050, South Africa \\\email{unnikrishnan.sureshkumar@wits.ac.za} 
    \label{aff:wits}
    \and
    Astronomical Observatory of the Jagiellonian University, ul. Orla 171, 30-244 Krak\'{o}w, Poland
    \label{aff:oauj}
    \and
    National Centre for Nuclear Research, ul. Pasteura 7, 02-093 Warsaw, Poland 
    \label{aff:ncbj}
    \and
    Centre of Excellence for Data Science, Artificial Intelligence and Modelling (DAIM), University of Hull, Cottingham Road, Hull, HU6 7RX, UK
    \label{aff:daim}
    \and
    E. A. Milne Centre for Astrophysics, University of Hull, Cottingham Road, Hull, HU6 7RX, UK
    \label{aff:eamilne}
    \and
    Department of Earth and Space Sciences, Indian Institute of Space Science \& Technology, Thiruvananthapuram 695547, Kerala, India
    \label{aff:iist}
    \and
    Astronomy Centre, University of Sussex, Falmer, Brighton BN1 9QH, UK
    \label{aff:sussex}
    \and
    Centre for Astrophysics and Supercomputing, Swinburne University of Technology, John Street, Hawthorn, 3122, Australia
    \label{aff:cas_swisburne}
    }

\date{Received 11 August 2023 / Accepted 28 February 2024}

\abstract
{
Galaxy mergers play a crucial role in galaxy evolution. 
However, the correlation between mergers and the local environment of galaxies is not fully understood.
}
{
We aim to address the question of whether galaxy mergers prefer denser or less dense environments by quantifying the spatial clustering of mergers and non-mergers.
We use two different indicators to classify mergers and non-mergers -- classification based on a deep learning technique ($f$) and  non-parametric measures of galaxy morphology, Gini-$M_{20}$ ($g$).
}
{
We used a set of galaxy samples in the redshift range $0.1 < z < 0.15$ from the Galaxy and Mass Assembly (GAMA) survey with a stellar mass cut of $\log \left(M_{\star}/\mathrm{M}_{\sun} \right) > 9.5$. 
We measured and compared the two-point correlation function (2pCF) of the mergers and non-mergers classified using the two merger indicators $f$ and $g$.
We measured the marked correlation function (MCF), in which the galaxies were weighted by $f$ to probe the environmental dependence of galaxy mergers.
}
{
We do not observe a statistically significant difference between the clustering strengths of mergers and non-mergers obtained using 2pCF.
However, using the MCF measurements with $f$ as a mark, we observe an anti-correlation between the likelihood of a galaxy being a merger and its environment.
Our results emphasise the advantage of MCF over 2pCF in probing the environmental correlations. 
}
{
Based on the MCF measurements, we conclude that the galaxy mergers prefer to occur in the under-dense environments on scales $> 50 \, h^{-1} \mathrm{kpc}$ of the large-scale structure (LSS).
We attribute this observation to the high relative velocities of galaxies in the densest environments that prevent them from merging.
}


\titlerunning{Do galaxy mergers prefer under-dense environments?}
\authorrunning{U. Sureshkumar et al.}

\maketitle

\section{Introduction}

In the hierarchical structure formation scenario, gas cools down at the centre of dark matter haloes to form galaxies \citep{press_schechter_1974, white&rees1978}.
These galaxies undergo various processes, of which galaxy mergers play an important role in galaxy evolution.
Galaxy mergers occur when two or more galaxies collide into each other, and these events affect the properties of the galaxies involved. 
For example, galaxy mergers can lead to the transformation of disk galaxies to elliptical ones \citep[e.g.][]{toomre&toomre1972, schweizer1982, barnes1992} and are expected to influence the star formation activity \citep[e.g.][]{larson1978, barnes1991, mihos1996, ellison2008, patton2013, davies2015, silva2018, pearson2019_merger_sfr, pearson2022}.
The galaxy mergers can also contribute towards the growth of supermassive black holes at the galaxy centres \citep[e.g.][]{volonteri2003, dotti2012, ellison2019}. 
All these effects make galaxy mergers an important phenomenon in the process of galaxy evolution. 

The environment, that is, the local density of galaxies, is expected to be a decisive factor in shaping the evolution of galaxies. 
However, what remains unclear is the exact role that the environment plays on the merger rates and the post-merger properties of the merging galaxies such as their morphology, star formation rate (SFR), stellar mass, luminosity, and the evolution of their stellar population. 
The primary question is whether galaxy mergers predominantly take place in the denser or less dense environments of the large-scale structure (LSS).

As the merging is an assembly of two or more galaxies into one, mergers are expected to have a strong dependence on the environment.
That is, we naively expect mergers to occur frequently in denser regions.
Some studies have tried to answer this question.
Indeed, a strong dependence of the galaxy merger rate on the environment was observed in high redshifts ($z \sim 1$), where higher merger rates were found in high density regions \citep{lin2010, deravel2011}.
However, only a weak dependence was found in the local Universe \citep{darg2010_gz_prop, ellison2010}.
The galaxies in massive cluster environments show less evidence of galaxy interactions and mergers than galaxies in groups or fields \citep{vanDokkum1999, mcintosh2008, tran2008, alonso2012, kampczyk2013}.
The relative velocities of galaxies in the heavily dense cluster environments are too high for the galaxies to merge with one another \citep{ostriker1980, mihos2004, deger2018, benavides2020}.  
\citet{skibba2009_morphology} observed that the probability of a galaxy to be undergoing a merger or interaction is almost uncorrelated with the environment except at projected separation scales of $100 \, h^{-1} \mathrm{kpc}$.
\citet{perez2009} found that intermediate density environments are efficient in close encounters and mergers.
\citet{delahaye2017} found no evidence of increased merger events in the cluster environment with respect to the field. 
\citet{lin2010} tried to study this issue by probing the environment of wet mergers (mergers between gas-rich, blue galaxies) and dry mergers (mergers between gas-poor, red galaxies).
They found that wet mergers mostly occur in low-dense environments and dry mergers occur in denser environments.

Simulations also indicate an environmental dependence of mergers, with different semi-analytic models showing varying strengths of the environmental dependence \citep{jian2012}.
It was observed in simulations as well that bound galaxy pairs tend to avoid extreme environments such as clusters and voids \citep{tonnesen2012}.
There have been investigations on the relation between halo merger rates and underlying environments using N-body simulations \citep{fakhouri2009, hester2010} resulting in contradicting results.
\citet{fakhouri2009} derived the merger rate of dark matter haloes as a function of local density and found that in regions with the highest density, galaxy-mass haloes experience mergers approximately 2.5 times more frequently than in other regions.
\citet{hester2010} conducted a similar study using subhaloes from simulations and found that the merger rate is roughly independent of the environment. 

In most of the above-mentioned works, the environment was parameterised using a local density parameter defined at a particular scale, either using nearest neighbours or fixed aperture \citep{muldrew2012}. 
Recently, \citet{omori2023} observed significant environmental correlations of galaxy mergers in both observations and simulations. 
They noted that the trend of these correlations varies depending on the scale at which the environmental parameter is defined.
Therefore, it is essential to study environmental correlations on a wide range of scales.
This approach helps in the investigation of environmental effects that are prevalent at different scales as well as minimising the impact of the chosen density parameter.
In this work, we implement this approach by using the galaxy two-point correlation function \citep[2pCF;][]{peebles1980} and the marked correlation function \citep[MCF;][]{stoyan&stoyan1994, beisbart2000}.

The 2pCF is useful for quantifying the strength of the spatial clustering of galaxies in a sample.
The amplitude of 2pCF strongly depends on various galaxy properties such as luminosities in optical \citep{zehavi2011, farrow2015_gama_cf, pollo2006}, ultraviolet \citep{heinis2004, milliard2007}, and infrared \citep{torre2007, solarz2015, sureshkumar2023} bands; stellar mass \citep{skibba2015, durkalec2018}; SFR \citep{hartley2010, lin2012, mostek2013}; colour \citep{coil2008, coupon2012, skibba2014_combining_fields}; and spectral type \citep{norberg2002, meneux2006}.
In general, these works conclude that galaxies that are more luminous, massive, evolved, and redder prefer to exist in the denser regions of the LSS than their counterparts.

On the other hand, the MCF is a useful tool for studying the environmental dependence of galaxy properties as a function of the separation scale of galaxy pairs \citep{sheth&tormen2004, sheth2005_galform_models, skibba2013}. 
In MCF analysis, we define a mark to all the galaxies based on the property that we are interested in.
This mark is then used to weight the galaxies during the clustering measurement (see Sect.~\ref{sec:measurement_mcf} for details). 
This method allows us to quantify the environmental trends associated with a given galaxy property as a function of separation scale of galaxy pairs.
Many works have used the MCF to study the environmental dependence of luminosity, stellar mass, colour, age, morphology, and SFR \citep{skibba2006, skibba2009_morphology, sheth2006, gunawardhana2018, sureshkumar2021, sureshkumar2023}.
It also demonstrates the subtle differences between the environmental dependence of different galaxy properties, which may not be strongly evident in 2pCF measurements.
In our previous works, we used MCF to study the differences between environmental correlations of stellar mass, SFR, luminosities in optical to near-infrared bands \citep{sureshkumar2021}, and mid-infrared bands \citep{sureshkumar2023} by using these properties as the mark.
In this work, we compare the 2pCF measurements between merger and non-merger samples defined based on different merger indicators (Sect.~\ref{sec:result_2pcf}).
We also compute MCFs with a merger indicator as marks to trace the environment of galaxy mergers (Sect.~\ref{sec:result_mcf}) in the Galaxy and Mass Assembly \citep[GAMA;][]{driver2009_gama_gen} survey. 

There are various ways to mark a galaxy as a merger. 
Visual identification is one way, but it is very time-consuming and subjective. 
Non-parametric morphological parameters such as concentration-asymmetry-smoothness (CAS) and Gini$-M_{20}$ have also been used to quantitatively identify merging galaxies \citep{abraham1996, conselice2000, conselice2003, lotz2004, lotz2008_gm20_def}. 
Merging galaxies can also be identified using their projected separation on the sky and the relative velocity \citep{zepf&koo1989, lefevre2000, ellison2008, duncan2019}.
Machine learning based techniques are also being implemented for merger classification \citep[e.g.][]{walmsley2019, pearson2019_identifying, pearson2019_merger_sfr, ferreira2020, pearson2022, suelves2023}.
In this work, we consider two parameters as indicators of galaxy mergers.
The first one is a deep learning based quantity $f$ which is defined as the probability of the galaxy to be a merger \citep{pearson2019_merger_sfr}.
The second one is $g$ which is based on the Gini and $M_{20}$ parameters \citep{lotz2004}. 
These parameters are explained in Sect.~\ref{sec:data_mergindicators}.

The paper is organised as follows. 
In Sect.~\ref{sec:data} we describe the general properties of the GAMA survey, the merger indicators, the sample selection methods and the random catalogue that are used in this work. 
Section~\ref{sec:measurements} explains the techniques used to measure the 2pCF and MCF, as well as the associated methodologies. 
The results are presented in Sect.~\ref{sec:results} and discussed in Sect.~\ref{sec:discussion}.
Finally, we conclude in Sect.~\ref{sec:conclusions}.

Throughout the paper, a flat $\Lambda$CDM cosmological model with $\Omega_{\text{M}}=0.3$ and $\Omega_\Lambda=0.7$ is adopted and the Hubble constant is parameterised via $h=H_0/100 \rm \, km \, s^{-1} \, Mpc^{-1}$. 
The distances are expressed in comoving coordinates and are in units of $h^{-1} \mathrm{Mpc}$.

\section{Data}\label{sec:data}

\subsection{Galaxy and Mass Assembly survey}\label{sec:data_gama}

Galaxy and Mass Assembly (GAMA) is a multi-wavelength spectroscopic survey that  provides a sampling from the UV to far-IR range of wavelengths (0.15~--~500~$\mu\mathrm{m}$) through 21 broad-band filters \citep{driver2009_gama_gen}.
Spanning across five regions on sky (three equatorial regions named G09, G12, and G15 and two southern regions G02 and G23), it observes galaxies down to an $r$-band flux limit of $r < 19.8$.
For a detailed description of the GAMA survey, we refer the reader to \citet{driver2009_gama_gen}, \citet{robotham2010_gama_tiling}, \citet{driver2011_gama_coredata}, and \citet{liske2015_gama_dr2}.

In this work, we used galaxies from the equatorial regions (G09, G12, and G15) of GAMA II data release 4\footnote{\url{http://www.gama-survey.org/dr4/}} \citep{driver2022_gamadr4}.
Each of these regions covers 12 × 5 $\text{deg}^2$ of the sky and has high spatial completeness and an overall redshift completeness of $\sim98\%$ with $z_\text{median} = 0.2$.
We took the galaxy positions and redshifts from \textsc{TilingCatv46} data management unit (DMU).
We selected only those galaxies that satisfy the conditions: (a)~\texttt{SURVEY\_CLASS} $\geq$ 4, (b)~\texttt{nQ} $\geq$  3, and (c)~\texttt{VIS\_CLASS} = 0, \texttt{VIS\_CLASS} = 1, or \texttt{VIS\_CLASS} = 255.
These filters are applied in order to select the GAMA main survey galaxies with a secure redshift flag which assures that the redshift has $>90\%$ chance of being correct.
The filters also avoid sources that are visually classified to be blends of stars or parts of other galaxies \citep{baldry2010}.

We used stellar mass of the galaxies to define the samples for our analysis.
We chose stellar mass over absolute magnitudes to define samples because stellar mass is expected to be the stronger tracer of halo mass than magnitudes and the halo mass is the most prominent driver of galaxy clustering \citep[e.g.][]{wechsler2018}.
We assigned stellar masses for the selected galaxies from \textsc{StellarMassesLambdarv24} DMU  \citep{taylor2011_gama_stellar_mass, wright2016}.
These stellar masses are estimated using the methods of \citet{taylor2011_gama_stellar_mass} applied to the \textsc{lambdar} photometry of \citet{wright2016}.
These methods are based on the stellar population synthesis modelling of broadband photometry using \citet{bruzual2003}  stellar evolution models, \citet{chabrier2003} initial mass function, and \citet{calzetti2000} dust law.
Since the stellar mass estimates are based on aperture-matched photometry, it is essential to scale them to account for flux lying beyond the finite aperture. 
For that, we applied a \texttt{fluxscale} correction to the stellar mass as described in \citet{taylor2011_gama_stellar_mass}.
The \texttt{fluxscale} value is a ratio between the $r$ band aperture flux and the total S\'{e}rsic flux.
We took the \texttt{fluxscale} values from \textsc{StellarMassesLambdarv20} DMU and applied the \texttt{fluxscale}-correction to all the stellar mass values of the galaxies in this work.
We retained only those galaxies with $0 < \texttt{fluxscale} < 5$. 

\begin{figure}[t]
    \centering
    \includegraphics[width=\linewidth]{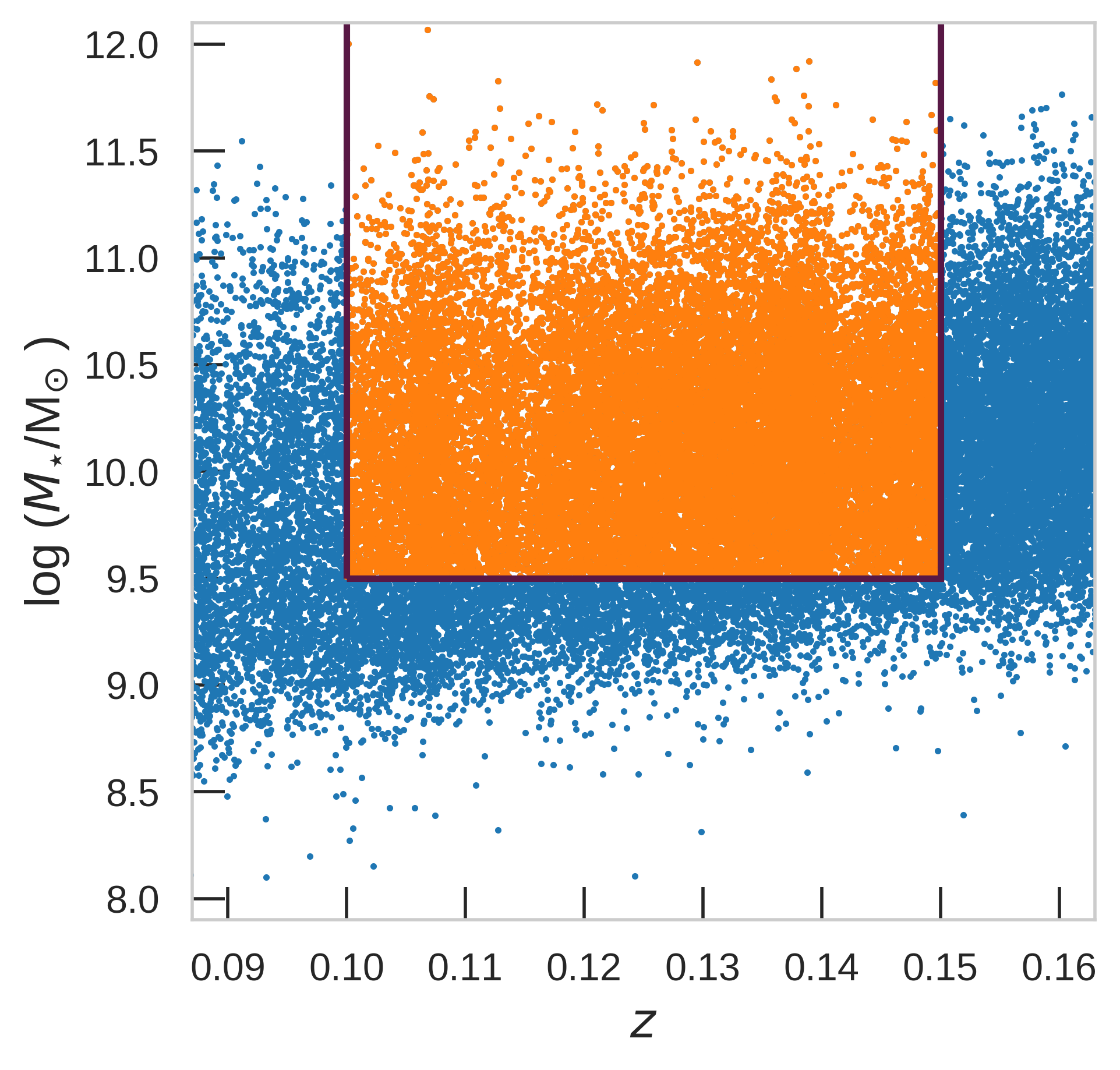}
    \caption{Redshift-stellar mass distribution of the GAMA galaxies selected for this work.
    The orange dots inside the rectangle show the main sample.}
    \label{fig:z-mass}
\end{figure}

We then selected a sample by applying a redshift cut of $0.1 < z < 0.15$ and stellar mass cut of $\log \left(M_{\star}/\mathrm{M}_{\sun} \right) > 9.5$.
Further, we retained those galaxies that are present in the merger classification catalogue of \citet{pearson2019_merger_sfr}.
This selection leads to a sample with 23855 galaxies which forms the main sample for the rest of this work.  
The redshift-stellar mass distribution of the main sample is represented by orange dots in Fig.~\ref{fig:z-mass}.
We divided the main sample into mergers and non-mergers using the merger indicators explained in the following section. 
The details of those samples are given in Sect.~\ref{sec:data_sampleselection}.

\subsection{Merger indicators}\label{sec:data_mergindicators}

We consider $f$ and $g$ as indicators of galaxy mergers. 
We use these two parameters to divide the main sample into mergers and non-mergers as described in the following sections.

\subsubsection{$f$ parameter}\label{sec:fracmerg}

\begin{figure*}[t]
    \centering
    \includegraphics[width=\linewidth]{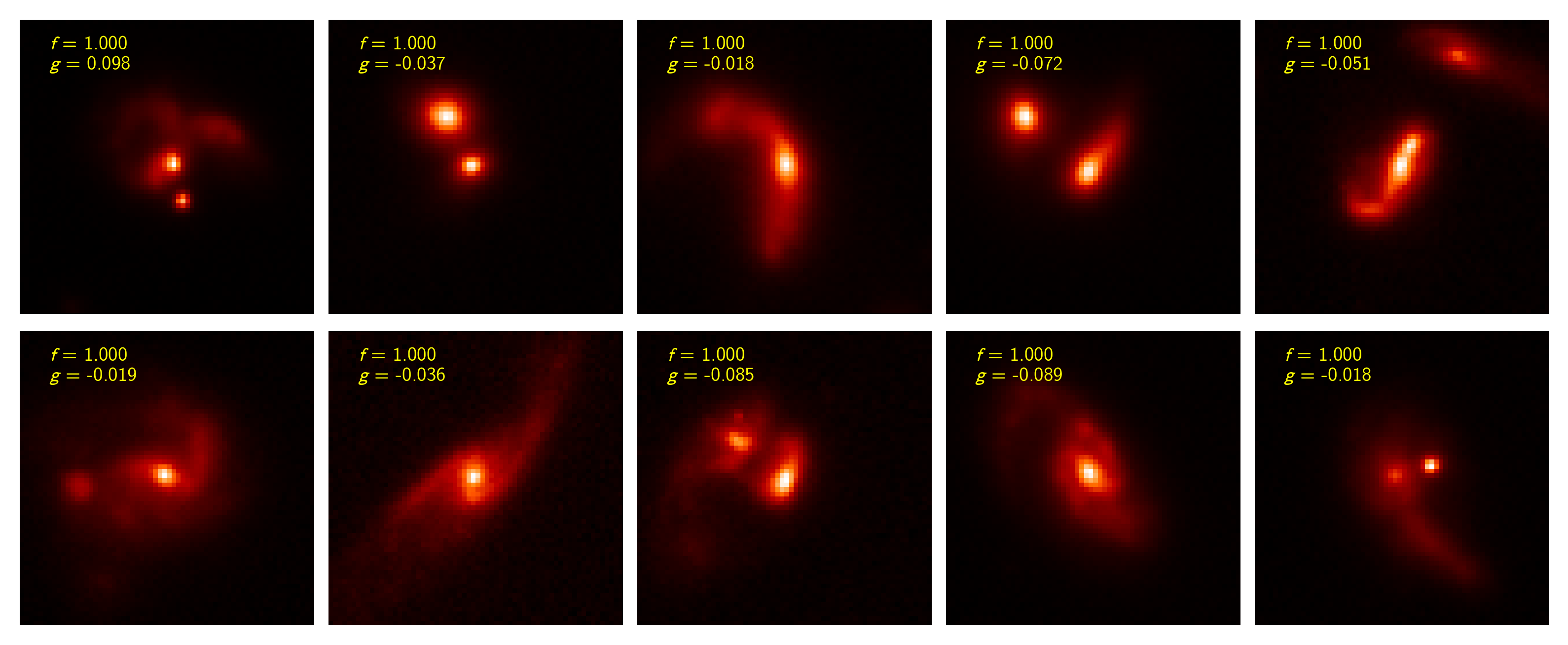}
    \caption{Optical images of ten objects in the main sample with the highest value of $f$. 
    The galaxies with $f > 0.52$ are more likely to be mergers.}
    \label{fig:cutout_fM}
\end{figure*}

The parameter $f$ is a probability that a galaxy is a merger, according to the convolutional neural network \citep[CNN;][]{lecun2015} used in \citet{pearson2019_merger_sfr}. 
This CNN was trained on $r$-band data from the Kilo Degree Survey \citep[KiDS;][]{dejong2013_kids}.
We refer the reader to \citet{pearson2019_merger_sfr} for a detailed description of the CNN. 
In short, the network takes one channel images of 64 $\times$ 64 pixels as an input. 
The images are scaled between 0 and 1. 
The network outputs two categories, merger and non-merger, with each parameter being a probability ranging between 0 and 1. 
The two parameters sum to unity. 
The output for the merger class is referred to as \texttt{frac\_merger} and we refer to and use this output as $f$ in this work to classify mergers. 
Galaxies with $f > 0.52$ are defined as mergers \citep{pearson2019_merger_sfr}. 
In Fig.~\ref{fig:cutout_fM} we show the optical images of ten objects from the main sample with the highest $f$ values.

\subsubsection{$g$ parameter}\label{sec:ginim20}

The $g$ parameter is derived from Gini and $M_{20}$ which are non-parametric measures of galaxy morphology.
The Gini coefficient $G$ describes how the light of a galaxy is distributed among the pixels \citep{lotz2004}. 
For a set of $n$ pixels with the $i$th pixel having the flux $f_i$, the Gini coefficient is defined as
\begin{equation}
    G = \frac{1}{\overline{|f|} n (n-1)} \sum\limits_{i}^{n} \left( 2i - n - 1 \right) |f_i| ,
\end{equation}
where $\overline{|f|}$ is the mean of the absolute flux over the pixel values.
$G=0$ implies that the light is distributed equally among all of the pixels and $G=1$ implies that all of the flux is concentrated in a single pixel.

The quantity $M_{20}$ is the second moment of the fluxes of the brightest pixels that have 20\% of the galaxy's light, normalised by the total second moment of all the pixels \citep{lotz2004}. 
The total second moment $M_\text{tot}$ is defined as the summation of the flux of a pixel multiplied by the squared distance of that pixel from the centre. 
That is,
\begin{equation}
    M_\text{tot} = \sum\limits_{i}^{n} M_i = \sum\limits_{i}^{n} f_i \left[ (x_i - x_\text{c})^2 + (y_i - y_\text{c})^2 \right] ,
\end{equation}
where $(x_i, y_i)$ is the two-dimensional spatial coordinates of the $i$th pixel and $(x_\text{c}, y_\text{c})$ is that of the galaxy centre.
The galaxy centre is defined using $(x_\text{c}, y_\text{c})$ that minimises the $M_\text{tot}$.

Then, $M_{20}$ is defined as,
\begin{equation}
    M_{20} = \log \left( \frac{\sum\limits_{i} M_i}{M_\text{tot}} \right), \, \text{while} \, \sum\limits_{i} f_i < 0.2 f_\text{tot} ,
\end{equation}
where $f_\text{tot}$ is the total flux of the pixels.

\begin{figure*}[t]
    \centering
    \includegraphics[width=\linewidth]{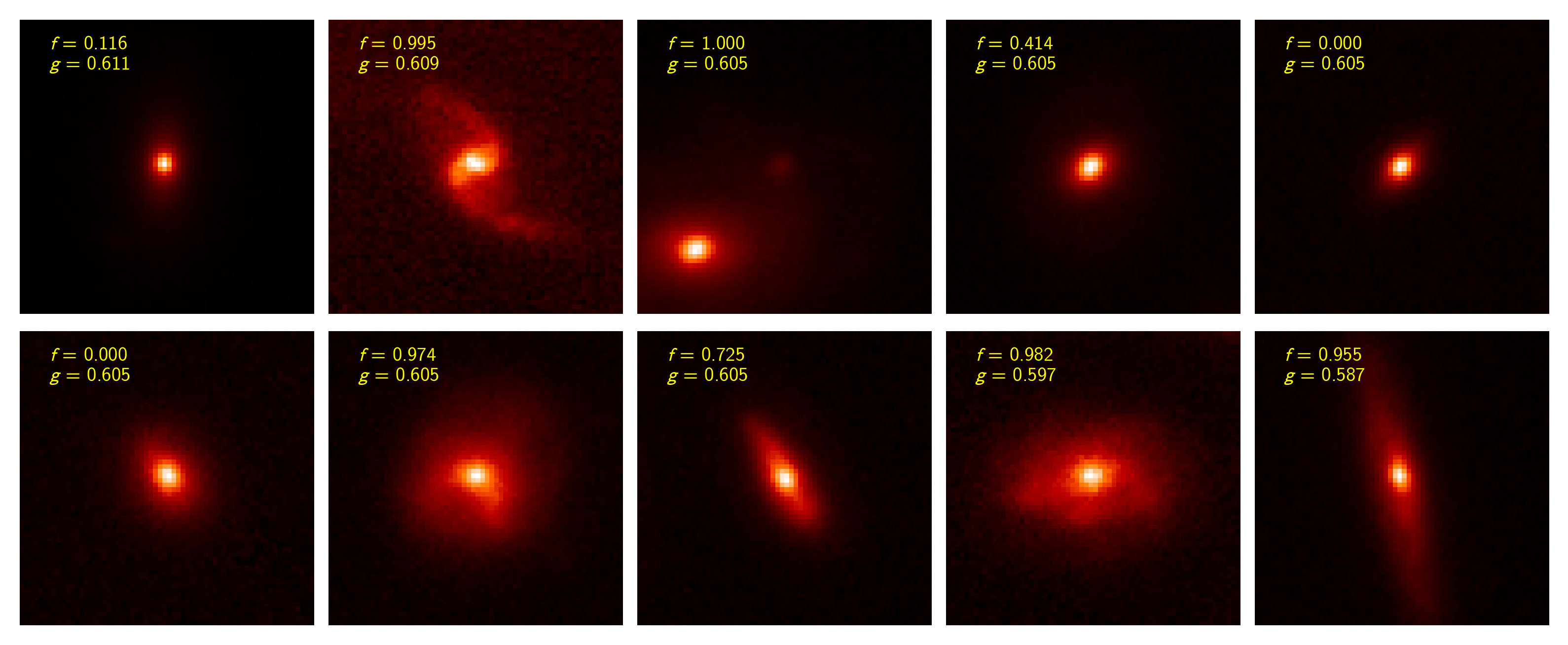}
    \caption{Optical images of ten objects in the main sample with the highest value of the $g$ parameter.
    The galaxies with $g > 0$ are more likely to be mergers.}
    \label{fig:cutout_gM}
\end{figure*}

Low redshift galaxies ($z<1.2$) are classified as mergers when
\begin{equation}
    G > -0.14 M_{20} + 0.33 ,
\end{equation}
as defined in \citet{snyder2015_morphology} and \citet{rodriguez-gomez2019}. 
Related to that, the Gini-$M_{20}$ merger statistics \texttt{gini\_m20\_merger} is defined as the perpendicular distance of a galaxy in the Gini-$M_{20}$ plane from the line that divides merging and non-merging galaxies \citep{snyder2015_structure, snyder2015_morphology, rodriguez-gomez2019}.
Based on this definition, mergers have $\texttt{gini\_m20\_merger} > 0$.
In this work, we refer \texttt{gini\_m20\_merger} as $g$.
For each of the galaxies in the main sample, we calculated $g$ from $G$ and $M_{20}$ derived using the python code \texttt{statmorph} \citep{rodriguez-gomez2019, pearson2019_merger_sfr}.
We present the optical images of ten galaxies in the main sample with the highest value of $g$ in Fig.~\ref{fig:cutout_gM}.

\subsection{Choice of the mark}\label{sec:data_markchoice}

We use $f$ as mark to compute the MCF.
We do not consider $g$ for the MCF computation.
The reason is that a higher value of $g$ does not necessarily correspond to a higher probability of an object being a merger. 
This is evident from the optical images presented in Fig.~\ref{fig:cutout_gM}. 
The galaxies with the highest $g$ values in the main sample show less evidence of a merging event than the galaxies with highest $f$ values (Fig.~\ref{fig:cutout_fM}).
Therefore, $g$ is not a good choice to weight the correlation function.

Moreover, several studies found that the $g$ parameter is a poor choice for merger identification \citep{lotz2008_merger_timescale, pearson2019_merger_sfr}.
For example, \citet{lotz2008_merger_timescale} discovered that the Gini-$M_{20}$ morphology had limitations in identifying mergers during the intermediate stage between the first pass and final merger, as well as after the coalescence stage.
\citet{pearson2019_merger_sfr} populated a set of visually classified non-mergers from KiDS-GAMA Galaxy Zoo \citep{lintott2008_galaxy_zoo} and visually identified mergers by \citet{darg2010_gz_prop, darg2010_gz_sdss}.
They found the distribution of mergers and non-mergers inconsistent with the classification scheme proposed by \citet{lotz2008_gm20_def} based on Gini-$M_{20}$. 
Studies such as \citet{snyder2019} and \citet{rose2023} demonstrated the better performance of machine learning-based algorithms, particularly random forest classification over the non-parametric morphological diagnostics.

\subsection{Merger and non-merger galaxy samples}\label{sec:data_sampleselection}

The main sample contains 23855 objects. 
The $f$ parameter divides the main sample into 6086 mergers and 17769 non-mergers. 
We label the merger sample selected based on $f$ as $\mathcal{F}_\mathrm{M}$ and the corresponding non-merger sample as $\mathcal{F}_\mathrm{NM}$.
The $g$ parameter divides the main sample into 882 mergers (sample $\mathcal{G}_\mathrm{M}$) and 22973 non-mergers (sample $\mathcal{G}_\mathrm{NM}$). 

\begin{figure}[t]
    \centering
    \includegraphics[width=\linewidth]{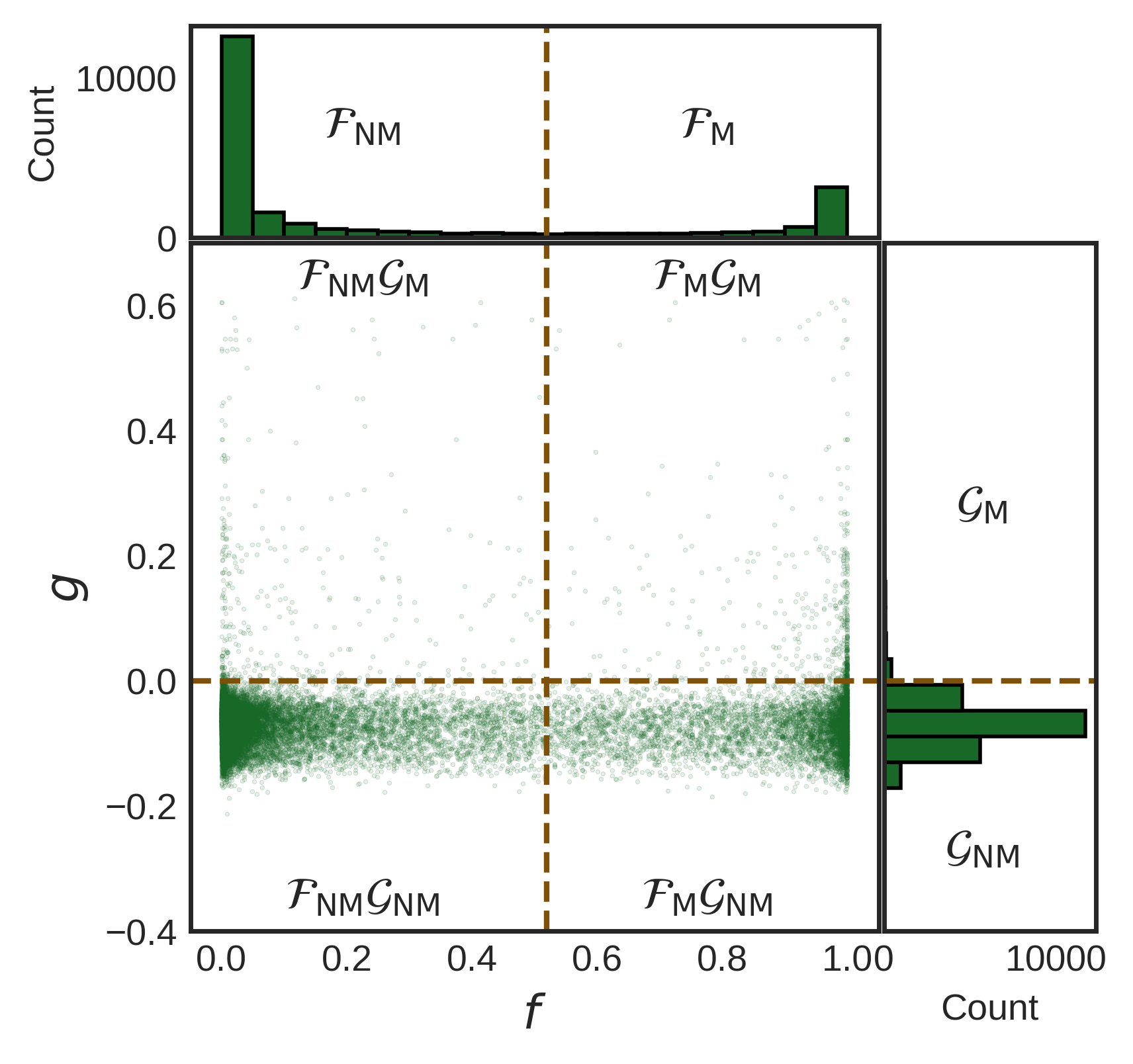}
    \caption{Distribution of $f$ and $g$ of the main sample.
    The dashed brown lines show the merger classification cuts $f = 0.52$ and $g = 0$.
    }
    \label{fig:frac_merg_hist}
\end{figure}

In Fig.~\ref{fig:frac_merg_hist}, we show the $g$--$f$ distribution of the main sample. 
The green dots refer to the 23855 main sample galaxies and dashed brown lines denote the merger classification cuts for the $f$ and $g$ parameters. 

\begin{figure*}[t]
    \centering
    \includegraphics[width=\linewidth]{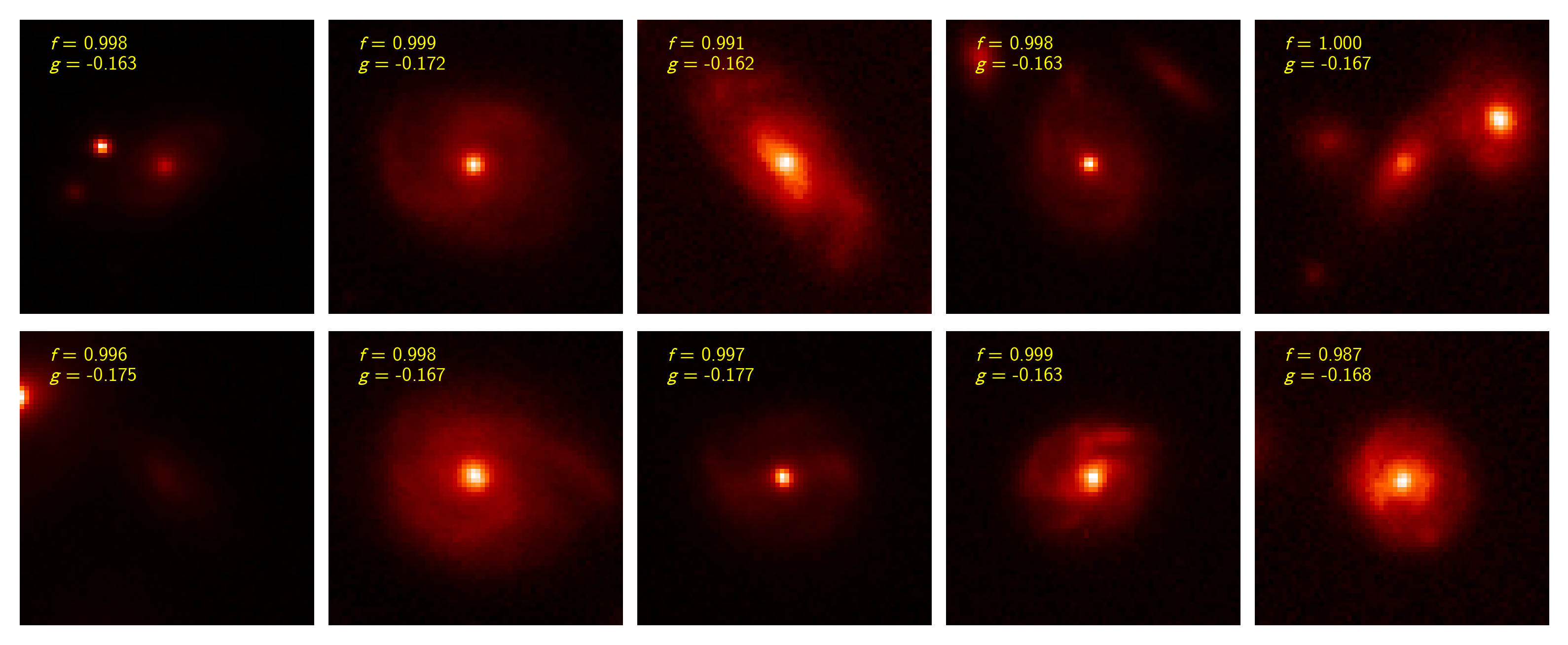}
    \caption{Representative examples from the sample $\mathcal{F}_\mathrm{M} \mathcal{G}_\mathrm{NM}$.
    These galaxies are taken from the bottom-right corner of Fig.~\ref{fig:frac_merg_hist} excluding those that are already shown in Fig.~\ref{fig:cutout_fM}.
    }
    \label{fig:cutout_fMgNM}
\end{figure*}

\begin{figure*}[t]
    \centering
    \includegraphics[width=\linewidth]{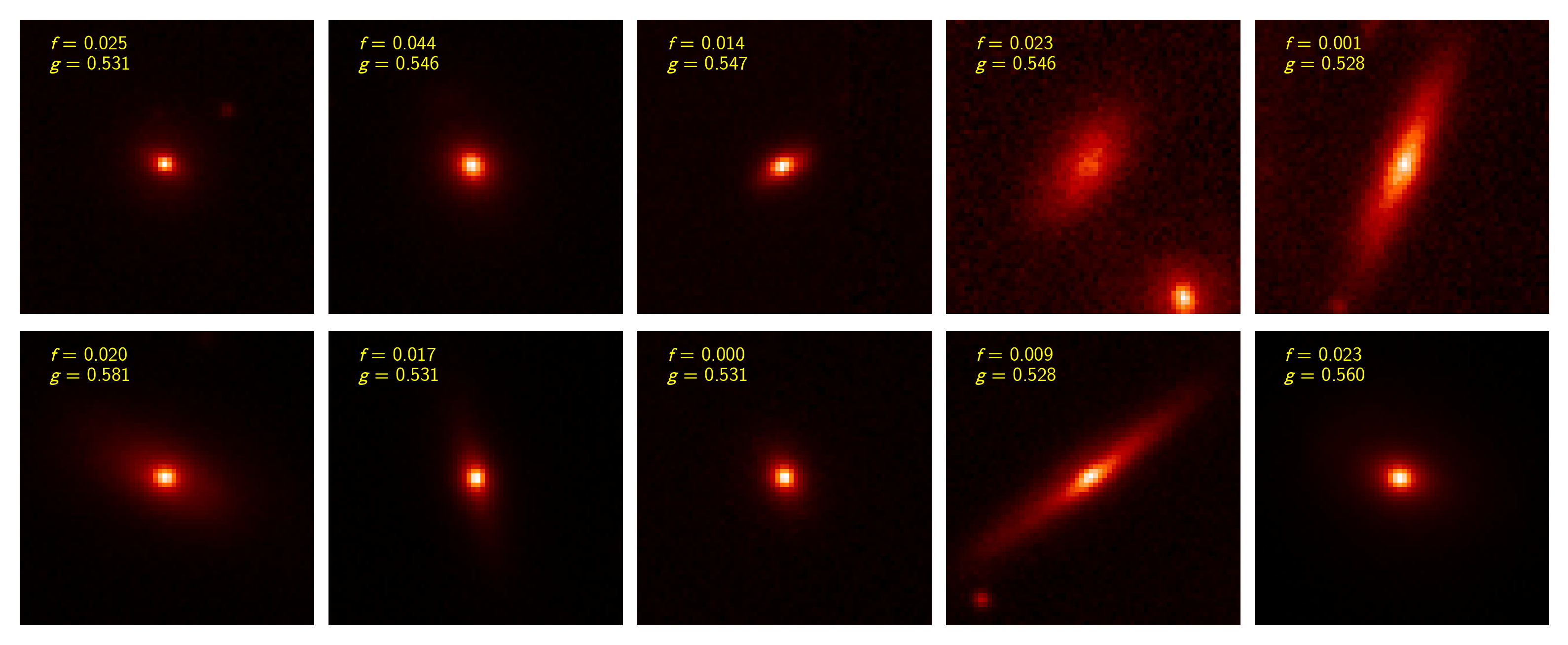}
    \caption{Representative examples from the sample $\mathcal{F}_\mathrm{NM} \mathcal{G}_\mathrm{M}$.
    These galaxies are taken from the top-left corner of Fig.~\ref{fig:frac_merg_hist} excluding those that are already shown in Fig.~\ref{fig:cutout_gM}.
    }
    \label{fig:cutout_fNMgM}
\end{figure*}

We also defined samples taking into account both the parameters together. 
The main sample contains 510 objects which are classified as mergers by both the parameters.
We label this sample as $\mathcal{F}_\mathrm{M} \mathcal{G}_\mathrm{M}$.
Similarly, we define sample $\mathcal{F}_\mathrm{M} \mathcal{G}_\mathrm{NM}$ with 5576 galaxies that are classified as mergers by $f$, but as non-mergers by $g$.
The sample $\mathcal{F}_\mathrm{NM} \mathcal{G}_\mathrm{M}$ contains 372 objects that are non-mergers based on $f$, but mergers based on $g$.
The remaining 17397 galaxies that are equally classified as non-mergers by both the parameters are labelled as $\mathcal{F}_\mathrm{NM} \mathcal{G}_\mathrm{NM}$.
Examples of galaxies from the samples $\mathcal{F}_\mathrm{M} \mathcal{G}_\mathrm{NM}$ and $\mathcal{F}_\mathrm{NM} \mathcal{G}_\mathrm{M}$ are shown in Fig.~\ref{fig:cutout_fMgNM} and Fig.~\ref{fig:cutout_fNMgM} respectively. 

The mean redshift of all the samples defined in this section is $z_\text{mean} = 0.13$.
The details of the samples such as the number of galaxies, number density, mean \textit{ugrJK}-absolute magnitudes, and mean stellar mass along with their standard deviations are given in Table~\ref{table:subsamples_properties}.

\begin{table*}[h]
\caption{Properties of merger and non-merger galaxy samples defined based on the $f$ and $g$ parameters.}
\begin{center}
\resizebox{\textwidth}{!}{
\begin{tabular}{l c c c c c c c c}
\toprule
\toprule
  \multicolumn{1}{l}{Sample} &
  \multicolumn{1}{c}{$N_\text{gal}$} &
  \multicolumn{1}{c}{\begin{tabular}[c]{@{}c@{}} $n_\text{gal}$\\ $({h}^{3} \, \text{Mpc}^{-3})$\end{tabular}} &
  \multicolumn{1}{c}{$M_u$} &
  \multicolumn{1}{c}{$M_g$} &
  \multicolumn{1}{c}{$M_r$} &
  \multicolumn{1}{c}{$M_J$} &
  \multicolumn{1}{c}{$M_K$} &
  \multicolumn{1}{c}{$\log \left(M_{\star}/\mathrm{M}_{\sun} \right)$} \\ 
  \midrule
  Main sample & 23855 & 0.0230 & $-18.75 \pm 0.85$ & $-20.05 \pm 0.85$ & $-20.66 \pm 0.89$ & $-21.43 \pm 0.98$ & $-21.40 \pm 1.03$ & $10.20 \pm 0.44$ \\ 
\midrule 
  $\mathcal{F}_\mathrm{M}$ & 6086 & 0.0059 & $-19.15 \pm 0.88$ & $-20.38 \pm 0.87$ & $-20.95 \pm 0.90$ & $-21.68 \pm 0.99$ & $-21.64 \pm 1.05$ & $10.27 \pm 0.45$ \\
 $\mathcal{F}_\mathrm{NM}$ & 17769 & 0.0172 & $-18.61 \pm 0.79$ & $-19.94 \pm 0.82$ & $-20.56 \pm 0.86$ & $-21.34 \pm 0.96$ & $-21.31 \pm 1.01$ & $10.17 \pm 0.43$ \\
  \midrule
   $\mathcal{G}_\mathrm{M}$ & 882 & 0.0009 & $-18.88 \pm 0.95$ & $-20.17 \pm 0.95$ & $-20.78 \pm 0.97$ & $-21.56 \pm 1.05$ & $-21.54 \pm 1.10$ & $10.27 \pm 0.48$ \\
  $\mathcal{G}_\mathrm{NM}$ & 22973 & 0.0222 & $-18.75 \pm 0.84$ & $-20.05 \pm 0.85$ & $-20.66 \pm 0.88$ & $-21.42 \pm 0.97$ & $-21.39 \pm 1.03$ & $10.20 \pm 0.44$ \\
  \midrule
  $\mathcal{F}_\mathrm{M} \mathcal{G}_\mathrm{M}$ & 510 & 0.0005 & $-19.09 \pm 0.97$ & $-20.36 \pm 0.96$ & $-20.96 \pm 0.99$ & $-21.74 \pm 1.07$ & $-21.71 \pm 1.13$ & $10.32 \pm 0.50$ \\
  $\mathcal{F}_\mathrm{M} \mathcal{G}_\mathrm{NM}$ & 5576 & 0.0054 & $-19.16 \pm 0.87$ & $-20.39 \pm 0.86$ & $-20.95 \pm 0.89$ & $-21.67 \pm 0.99$ & $-21.64 \pm 1.04$ & $10.26 \pm 0.45$ \\
  $\mathcal{F}_\mathrm{NM} \mathcal{G}_\mathrm{M}$ & 372 & 0.0004 & $-18.59 \pm 0.85$ & $-19.91 \pm 0.86$ & $-20.54 \pm 0.89$ & $-21.33 \pm 0.96$ & $-21.31 \pm 1.01$ & $10.19 \pm 0.44$ \\
  $\mathcal{F}_\mathrm{NM} \mathcal{G}_\mathrm{NM}$ & 17397 & 0.0168 & $-18.61 \pm 0.79$ & $-19.94 \pm 0.82$ & $-20.56 \pm 0.86$ & $-21.34 \pm 0.96$ & $-21.31 \pm 1.01$ & $10.17 \pm 0.43$ \\
\bottomrule
\end{tabular}
}
\end{center}
\tablefoot{
The columns represent the sample label, number of galaxies, number density, mean absolute magnitudes in the $u$, $g$, $r$, $J$, and $K$ bands, and mean stellar mass of the corresponding sample. 
The uncertainty with each property represents its standard deviation within the sample.
The sample definitions are as follows -- 
$\mathcal{F}_\mathrm{M}$: $f > 0.52$, 
$\mathcal{F}_\mathrm{NM}$: $f \leq 0.52$, 
$\mathcal{G}_\mathrm{M}$: $g > 0$, 
$\mathcal{G}_\mathrm{NM}$: $g \leq 0$, 
$\mathcal{F}_\mathrm{M} \mathcal{G}_\mathrm{M}$: $f > 0.52$ \& $g > 0$,
$\mathcal{F}_\mathrm{M} \mathcal{G}_\mathrm{NM}$: $f > 0.52$ \& $g \leq 0$,
$\mathcal{F}_\mathrm{NM} \mathcal{G}_\mathrm{M}$: $f \leq 0.52$ \& $g > 0$,
$\mathcal{F}_\mathrm{NM} \mathcal{G}_\mathrm{NM}$: $f \leq 0.52$ \& $g \leq 0$.
}
\label{table:subsamples_properties}
\end{table*}

\subsection{Random samples}\label{sec:data_random}

The 2pCF is a comparison of the spatial distribution of observed galaxies with respect to a random distribution.
Therefore, to measure the 2pCF of a galaxy sample (referred to as real sample), we need a random sample of galaxies with the same redshift and sky distribution.
We used the GAMA random galaxy catalogue (\textsc{Randomsv02} DMU) by \citet{farrow2015_gama_cf} to generate random galaxies corresponding to the selected sample.
The random catalogue is created using the method of \citet{cole2011} that accounts for targeting and redshift incompleteness of the GAMA survey.
The random galaxies are clones of the real galaxies and carry the same physical properties of the real galaxies.
We assigned the stellar mass, $f$, and $g$ values to each random galaxy by matching on the \texttt{CATAID} of the real galaxy.
For every real galaxy samples selected in Sect.~\ref{sec:data_sampleselection}, we applied the same redshift, stellar mass, and $f$ or $g$ selection cuts to the  random catalogue and then randomly select 10-20 times the number of real galaxies.

\section{Measurement methods}\label{sec:measurements}

In this work, we use galaxy 2pCF and MCF to study the dependence of mergers on environment. 
We refer the reader to the methodology sections of our previous works \citep{sureshkumar2021, sureshkumar2023} for a detailed description of the tools we use in this work, while we provide a short summary in this section.

\subsection{The galaxy two-point correlation function}\label{sec:measurement_2pcf}

The galaxy 2pCF, $\xi(r)$ is defined as the excess probability of observing a galaxy pair separated by $r$ over a random distribution \citep{peebles1980}.
To estimate the 2pCF at a separation scale $r$, we count the number of galaxy pairs that are separated by $r$. 
This counting process is performed separately for pairs formed from galaxies within the real observed sample (referred to as $DD(r)$), pairs from a randomly generated sample ($RR(r)$), and pairs formed between galaxies in the real and the random sample ($DR(r)$).

As a standard practice to account for the redshift-space distortion \citep{kaiser1987}, we estimated the two-dimensional 2pCF as a function of the components of pair separation that are perpendicular ($r_\text{p}$) and parallel ($\pi$) to the line-of-sight.
We used the \citet{landy&szalay1993} estimator given by
\begin{equation} \label{eqn:landy-szalay} 
    \xi(r_\text{p},\pi)= \frac{\langle DD(r_\text{p},\pi) \rangle - 2 \langle DR(r_\text{p},\pi) \rangle + \langle RR(r_\text{p},\pi) \rangle}{\langle RR(r_\text{p},\pi) \rangle} ,
\end{equation}
where $\langle DD \rangle$, $\langle DR \rangle$ and $\langle RR \rangle$ are the normalised real-real, real-random, and random-random pair counts respectively.

We then computed the projected 2pCF $\omega_\mathrm{p}(r_\mathrm{p})$ by integrating $\xi(r_\mathrm{p},\pi)$ along the line of sight ($\pi$) direction, given by
\begin{equation}\label{eqn:projectedcf}
    \omega_\mathrm{p}(r_\mathrm{p}) = 2 \, \int_0^{\pi_\text{max}} \xi(r_\mathrm{p},\pi) \, \mathrm d\pi .
\end{equation}

As in our previous works with GAMA data \citep{sureshkumar2021, sureshkumar2023}, we chose the limit of integration $\pi_\text{max}$ to be $40 \, h^{-1} \mathrm{Mpc}$ that is reasonable enough to include all the correlated galaxy pairs and reduce the noise in the estimator.

\subsubsection{Error estimate of two-point correlation function}\label{sec:measurement_errors_2pcf}

We used jackknife method of error estimation \citep{norberg2009}. 
The number of jackknife samples ($N_\text{JK}$) should be $>N^{3/2}_\text{bin}$ so that the covariance matrix is not too noisy \citep{mandelbaum2006_JKNr} where $N_\text{bin}$ is the number of separation bins in which the 2pCF is measured.
In our case, $N_\text{bin} = 8$ and hence we chose $N_\text{JK} = 24$.
The associated covariance matrix is given by
\begin{equation}\label{eqn:covmat_jackknife}
C_{ij} = \frac{N_\text{JK} - 1}{N_\text{JK}} \sum\limits_{k=1}^{N_\text{JK}} \left( \omega_\mathrm{p}^k(r_i) - \bar\omega_\mathrm{p}(r_i) \right) \, \, \left( \omega_\mathrm{p}^k(r_j) - \bar\omega_\mathrm{p}(r_j) \right) ,
\end{equation}
where $\omega_\mathrm{p}^k(r_j)$ is the CF value in $r_j$ bin of the $k$th jackknife copy and $\bar{\omega_\mathrm{p}}$ is the mean CF from $N_\text{JK}$ copies. 
The square roots of the diagonal elements of the covariance matrix define the error bars of $\omega_\mathrm{p}$.

As proposed by \citet{hartlap2007}, we corrected the inverse covariance matrix by $(N_\text{JK} - N_\text{bin} - 2)/(N_\text{JK} - 1)$.
The corrected inverse covariance matrix is further used for the modelling of 2pCF.

\subsubsection{Halo occupation distribution modelling of two-point correlation function}\label{sec:measurement_HODmodel}

To connect the galaxy clustering measurements with the environment, we modelled 2pCF within the framework of halo occupation distribution \citep[HOD; ][]{cooray2002} models. 
These models describe the probability $P(N|M_\text{h})$ of a dark matter halo of mass $M_\text{h}$ hosting $N$ galaxies of a given type. 

The halo occupation function $\langle N(M_\text{h}) \rangle$ is composed of the mean occupation functions for the central $\langle N_\text{cen}(M_\text{h}) \rangle$ and satellite $\langle N_\text{sat}(M_\text{h}) \rangle$ galaxies. 
Following common analytical prescription proposed by \cite{zheng2007_HOD} we parameterised the HOD model using five free parameters:
\begin{equation}
    \langle N(M_\text{h}) \rangle = \langle N_\text{cen}(M_\text{h}) \rangle + \langle N_\text{sat}(M_\text{h}) \rangle ,
\end{equation}
where
\begin{equation}
    \langle N_\text{cen} (M_\text{h}) \rangle
= \frac{1}{2} \left[ 1 + \text{erf} \left( \frac{\log M_\text{h} - \log M_\text{min}}{\sigma_{\log M}} \right) \right] ,
\end{equation}
\begin{equation}
    \langle N_\text{sat} (M_\text{h}) \rangle = \langle N_\text{cen} (M_\text{h}) \times \left( \frac{M_\text{h} - M_0}{M'_1} \right)^\alpha ,
\end{equation}
where $\text{erf}$ is the error function $\text{erf} (x) = \frac{2}{\sqrt{\pi}} \int_0^x e^{-t^2} \mathrm d t$.
$M_\text{min}$, $M_{1}$ and $M_0$ denote the minimum, satellite, and cut off halo masses respectively.
The scatter between the stellar mass of the galaxies and the halo mass is given by $\sigma_{\log M}$, while $\alpha$ is the power-law slope of the galaxy mean occupation function.

From the best-fitting HOD parameters we derived quantities describing the halo and galaxy properties, such as the average host halo mass $\langle M_\text{h} \rangle$,
\begin{equation}
    \langle M_\text{h} \rangle (z) = \int \mathrm d M_\text{h} \, M_\text{h} \, n(M_\text{h}, z) \, \frac{\langle N (M_\text{h}) \rangle}{n_\text{g} (z)} ,
\end{equation}
and the fraction of satellite galaxies per halo $f_s$,
\begin{equation}
    f_\text{s} = 1 - \int \mathrm d M_\text{h} \, n(M_\text{h}, z) \, \frac{\langle N_\text{cen} (M_\text{h})}{n_\text{g}(z)} ,
\end{equation}
where $n(M_\text{h}, z)$ is the dark matter mass function and $n_
\text{g}(z)$ is the number density of galaxies
\begin{equation}
    n_\text{g}(z) = \int \mathrm d M_\text{h} \, n(M_\mathrm{h}, z) \, \langle N_\text{g} (M_\text{h}) \rangle .
\end{equation}

We implemented a Markov chain Monte Carlo (MCMC) technique using the full covariance matrix (Eq.~\ref{eqn:covmat_jackknife}) to explore the HOD parameter space for all the merger and non-merger galaxy samples described in Sect.~\ref{sec:data_sampleselection}.
Following \citet{mccracken2015}, we computed the real-space correlation function $\xi(r)$ from the best-fitting HOD parameters.
Then we directly measured the correlation length $r_0$ as the value of $\xi$ at $r = 1 \, h^{-1} \text{Mpc}$. 
The uncertainty in correlation length for a given sample was estimated from the standard deviation of $r_0$ over the last 5000 MCMC iterations of halo model fits.  

Additionally, we also tested for the effect of integral constraint \citep[IC;][]{infante1994, roche1999} due to the limited volume of the GAMA survey. 
We computed the IC in the real-space 2pCF given by
\begin{equation}
    C_\xi = \frac{\sum_i \xi^\text{model}(r_i) RR(r_i)}{\sum_i RR(r_i)} ,
\end{equation}
where $RR(r_i)$ is the number of pairs in the random sample with separation in the linearly spaced $r_i$ bins and $\xi^\text{model}(r_i)$ is the modelled real-space 2pCF.
Then the corresponding IC in the projected 2pCF \citep{newman2024, clontz2022} is given by
\begin{equation}
    C_{\omega_\mathrm{p}} = 2 \, \pi_\text{max} \, C_\xi ,
\end{equation}
where $\pi_\text{max} = 40 \, h^{-1} \mathrm{Mpc}$.
We computed the IC at every 100th iteration of each 25 MCMC chains for each sample.
We find an average $C_{\omega_\mathrm{p}}$$\sim$$2 \, h^{-1} \mathrm{Mpc}$ for all the samples.
This is about 38-52\% of the error bars of projected CF at the largest scale ($r_\mathrm{p} \sim 13 \, h^{-1} \mathrm{Mpc}$). 
Therefore, we observe no significant effect of IC on our CF measurements. 

\subsection{Marked correlation function}\label{sec:measurement_mcf}

The marked correlation function is an efficient statistical tool to study the environmental dependence of galaxy properties \citep{sheth2005_galform_models}.
The measurement of MCF involves marking all the real galaxies with a physical property (called `mark') such as luminosity, colour, stellar mass, SFR, morphology, or, as in this work, probability of the galaxy to be a merger.
This mark is then used to weight the galaxies while counting the pairs in the separation bins. 
That is, the two-dimensional MCF in real space is defined as
\begin{equation}\label{eqn:M(r)}
M(r_\mathrm{p}, \pi) = \frac{1 + W(r_\mathrm{p}, \pi)}{1 + \xi(r_\mathrm{p}, \pi)} ,
\end{equation}
where $W(r_\mathrm{p}, \pi)$ is the weighted CF computed using the \citet{landy&szalay1993} estimator in Eq.~\ref{eqn:landy-szalay}, but with all the real galaxies weighted by the ratio of their mark to the mean mark of the sample.
That is, the weight of a real galaxy is given by $\text{weight} = \text{mark} / \overline{\text{mark}}$.

The MCF measurements can also be affected by the redshift-space distortions.
So we measured the projected MCF given by
\begin{equation}\label{eqn:projectedMCF}
    M_\mathrm{p}(r_\mathrm{p}) = \frac{1 + W_\mathrm{p}(r_\mathrm{p})/r_\mathrm{p}}{1 + \omega_\mathrm{p}(r_\mathrm{p})/r_\mathrm{p}} ,
\end{equation}
where $W_\mathrm{p}(r_\mathrm{p})$ is the projected weighted CF, computed by integrating $W(r_\mathrm{p}, \pi)$ along the line-of-sight.

The $\mathrm{MCF}=1$ for a given property as mark implies a lack of correlation between that property and the environment.
Stronger deviation of MCF above unity ($\mathrm{MCF} > 1$) means greater correlation with the environment and stronger deviation below unity ($\mathrm{MCF} < 1$) means greater anti-correlation between the property and the environment. 

To estimate the error bars of MCF, we combined the jackknife error and random shuffling error using summation in quadrature method.
The random shuffling error is obtained by randomly shuffling the marks among the galaxies in the sample and then remeasuring the MCF. 
The standard deviation over $\sim$ 100 such MCF measurements gives the uncertainty in the MCF \citep{beisbart2000, skibba2006}.

\section{Results}\label{sec:results}

In this section we present the 2pCF and MCF measurements in different samples of GAMA galaxies in the redshift range $0.1 < z < 0.15$ and with stellar mass limit $\log \left(M_{\star}/\mathrm{M}_{\sun} \right) > 9.5$. 
In Sect.~\ref{sec:result_2pcf} we present the 2pCF measurements and their best-fitting HOD parameters of all the samples defined in Sect.~\ref{sec:data_sampleselection}.
Section~\ref{sec:result_mcf} presents the MCF measurements of the main galaxy sample with $f$ as the mark.

\subsection{Two-point correlation functions}\label{sec:result_2pcf}

\begin{figure*}[t]
    \centering
    \includegraphics[width=\linewidth]{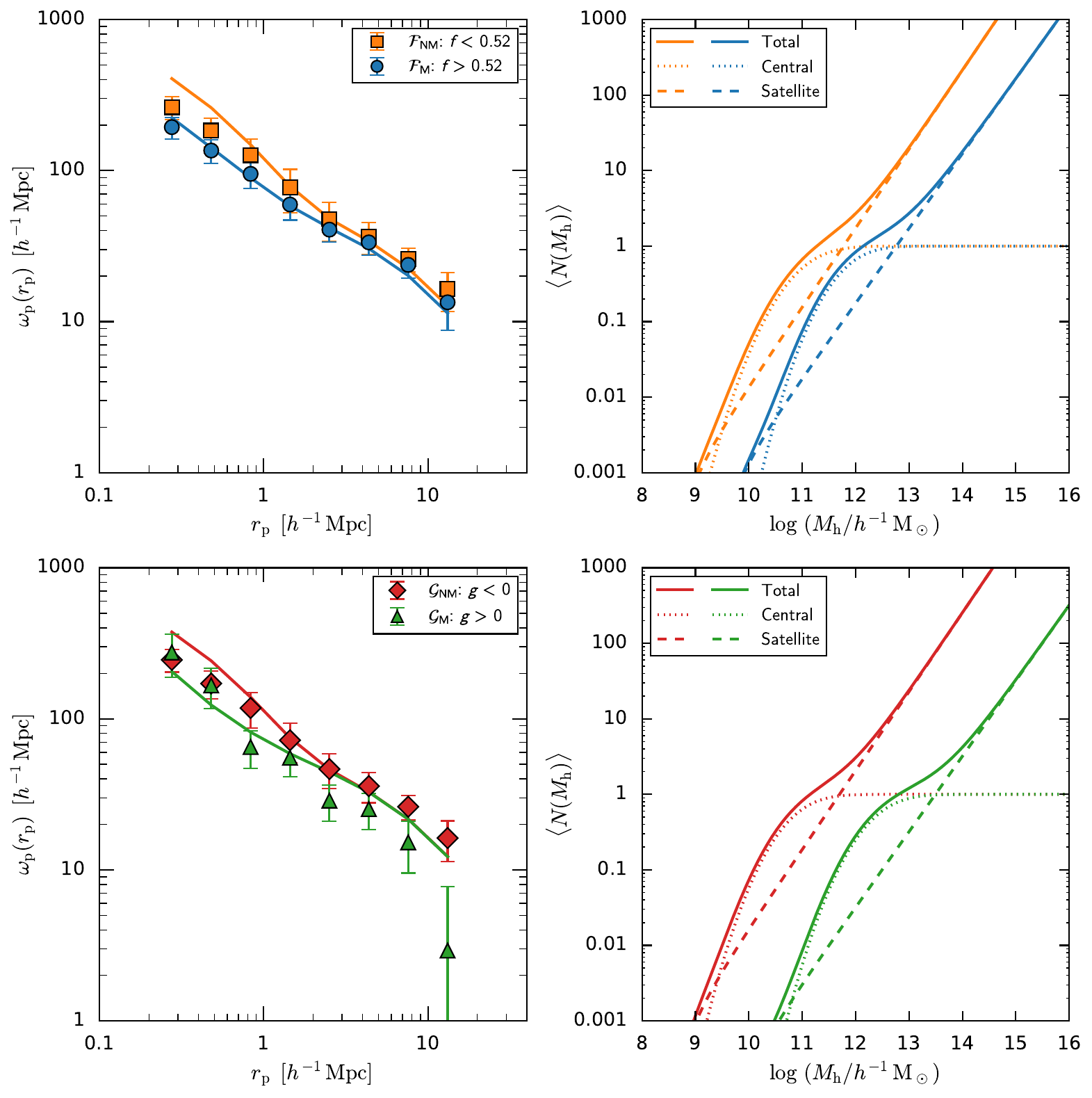}
    \caption{Projected 2pCFs of the mergers and non-mergers and the best-fitting HOD models (solid curves) when the division was performed based on $f$ (samples $\mathcal{F}_\mathrm{M}$ and $\mathcal{F}_\mathrm{NM}$: blue circles and orange squares, respectively, in the \textit{top left panel}) and $g$ (samples $\mathcal{G}_\mathrm{M}$ and $\mathcal{G}_\mathrm{NM}$: green triangles and red diamonds, respectively, in the \textit{bottom left panel}).
    The error bars on the markers are the square root of the diagonals of the covariance matrix obtained from the jackknife resampling method.
    The right panels show the corresponding total (solid), central (dotted), and satellite (dashed) halo occupation functions.
    }
    \label{fig:merger-nonmerger-wp}
\end{figure*}

\begin{figure*}[t]
    \centering
    \includegraphics[width=\linewidth]{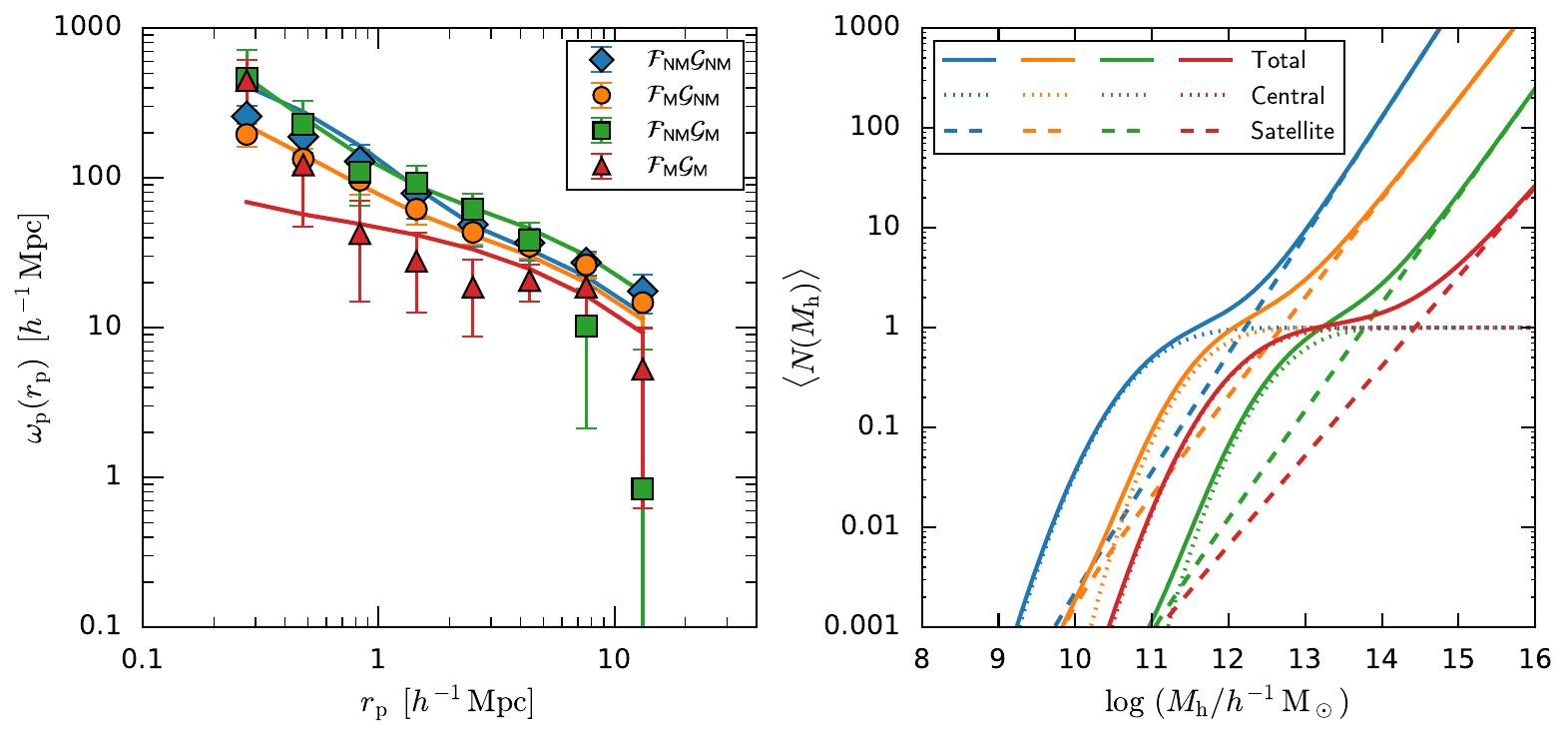}
    \caption{Projected 2pCFs with best-fitting HOD models (\textit{left panel}) of the samples $\mathcal{F}_\mathrm{NM} \mathcal{G}_\mathrm{NM}$ (blue diamonds), $\mathcal{F}_\mathrm{M} \mathcal{G}_\mathrm{NM}$ (orange circles), $\mathcal{F}_\mathrm{NM} \mathcal{G}_\mathrm{M}$ (green squares), and $\mathcal{F}_\mathrm{M} \mathcal{G}_\mathrm{M}$ (red triangles).
    The error bars on the markers are the square root of the diagonals of the covariance matrix obtained from the jackknife resampling method.
    The right panel shows the corresponding total (solid), central (dotted), and satellite (dashed) halo occupation functions.
    }
    \label{fig:merger-nonmerger-wp-cross}
\end{figure*}

In the top left panel of Fig.~\ref{fig:merger-nonmerger-wp} we present the projected 2pCF and the best-fitting HOD models for mergers and non-mergers for the classification cut $f = 0.52$.
In the bottom left panel we present the similar measurements for the classification cut $g = 0$.
The left panel of Fig.~\ref{fig:merger-nonmerger-wp-cross} shows the 2pCFs of samples with joined merger classification based on both parameters. 
All the 2pCFs were measured for the projected separation scales of $0.27 < r_\mathrm{p}/h^{-1} \mathrm{Mpc} < 13$.
As described in Sect.~\ref{sec:measurement_errors_2pcf}, we obtained the errors for 2pCFs from 24 jackknife realisations.
We fit the 2pCFs with HOD model using the method described in Sect.~\ref{sec:measurement_HODmodel}.
The corresponding halo occupation functions are shown in the right panels of Fig.~\ref{fig:merger-nonmerger-wp} and Fig.~\ref{fig:merger-nonmerger-wp-cross}.
In Table~\ref{table:wp_params} we present the best fitting HOD parameters along with the derived quantities of all the samples defined in Table~\ref{table:subsamples_properties}.

\begin{table*}[h]
\caption{Best-fitting HOD parameters and the derived correlation length of all samples defined in Table~\ref{table:subsamples_properties}.}
\begin{center}
\begin{tabular}{l c c c c c c c c c}
\toprule
\toprule
  \multicolumn{1}{l}{Sample} &
  \multicolumn{1}{c}{$\log M_\text{min}$} &
  \multicolumn{1}{c}{$\log M_\text{1}$} &
  \multicolumn{1}{c}{$\alpha$} &
  \multicolumn{1}{c}{$\sigma_{\log M}$} &
  \multicolumn{1}{c}{$\log M_\text{0}$} &
  \multicolumn{1}{c}{$\log \langle M_\text{h} \rangle$} &
  \multicolumn{1}{c}{$f_\text{s}$} &
  \multicolumn{1}{c}{$r_0$} &
  \multicolumn{1}{c}{$\chi^2/\text{dof}$}  \\[1.5mm]
  \midrule
$\mathcal{F}_\mathrm{M}$ & 11.79$^{+1.06}_{-0.52}$ & 12.76$^{+1.3}_{-0.54}$ & 0.99$^{+0.13}_{-0.22}$ & 0.71$^{+0.4}_{-0.44}$ & 9.46$^{+1.53}_{-1.46}$ & $13.28 \pm 0.06$ & $0.38 \pm 0.21$ & $6.11 \pm 0.55$ & 1.53 \\[1.5mm]
$\mathcal{F}_\mathrm{NM}$ & 10.98$^{+1.5}_{-0.73}$ & 11.77$^{+1.53}_{-0.83}$ & 1.05$^{+0.07}_{-0.09}$ & 0.78$^{+0.44}_{-0.45}$ & 8.59$^{+1.09}_{-1.54}$ & $13.47 \pm 0.03$ & $0.54 \pm 0.16$ & $6.62 \pm 0.30$ & 1.09 \\[1.5mm]
\midrule
$\mathcal{G}_\mathrm{M}$ & 12.36$^{+1.72}_{-0.8}$ & 13.49$^{+1.71}_{-1.21}$ & 1.0$^{+0.28}_{-0.44}$ & 0.76$^{+0.43}_{-0.46}$ & 9.75$^{+1.87}_{-1.87}$ & $13.22 \pm 0.31$ & $0.25 \pm 0.26$ & $6.43 \pm 1.07$ & 1.87 \\[1.5mm]
$\mathcal{G}_\mathrm{NM}$ & 10.82$^{+1.35}_{-0.74}$ & 11.7$^{+1.42}_{-0.8}$ & 1.04$^{+0.08}_{-0.1}$ & 0.74$^{+0.41}_{-0.46}$ & 8.57$^{+1.07}_{-1.48}$ & $13.45 \pm 0.03$ & $0.52 \pm 0.17$ & $6.44 \pm 0.29$ & 1.07 \\[1.5mm]
\midrule
$\mathcal{F}_\mathrm{M} \mathcal{G}_\mathrm{M}$ & 12.29$^{+1.72}_{-1.06}$ & 14.43$^{+1.71}_{-1.37}$ & 0.9$^{+0.5}_{-0.6}$ & 0.83$^{+0.48}_{-0.44}$ & 9.75$^{+1.85}_{-1.91}$ & $12.76 \pm 0.82$ & $0.04 \pm 0.31$ & $5.25 \pm 1.48$ & 1.85 \\[1.5mm]
$\mathcal{F}_\mathrm{M} \mathcal{G}_\mathrm{NM}$ & 11.72$^{+1.18}_{-0.55}$ & 12.69$^{+1.48}_{-0.67}$ & 0.99$^{+0.12}_{-0.2}$ & 0.69$^{+0.39}_{-0.46}$ & 9.39$^{+1.53}_{-1.44}$ & $13.28 \pm 0.14$ & $0.39 \pm 0.22$ & $6.08 \pm 0.65$ & 1.53 \\[1.5mm]
$\mathcal{F}_\mathrm{NM} \mathcal{G}_\mathrm{M}$ & 12.85$^{+1.52}_{-0.74}$ & 13.77$^{+1.57}_{-0.74}$ & 1.07$^{+0.32}_{-0.39}$ & 0.75$^{+0.42}_{-0.42}$ & 10.13$^{+2.12}_{-1.95}$ & $13.47 \pm 0.26$ & $0.28 \pm 0.25$ & $8.09 \pm 1.31$ & 2.12 \\[1.5mm]
$\mathcal{F}_\mathrm{NM} \mathcal{G}_\mathrm{NM}$ & 11.05$^{+1.25}_{-0.65}$ & 12.22$^{+1.28}_{-0.61}$ & 1.18$^{+0.1}_{-0.11}$ & 0.81$^{+0.5}_{-0.58}$ & 8.78$^{+1.21}_{-1.61}$ & $13.50 \pm 0.02$ & $0.34 \pm 0.11$ & $6.50 \pm 0.37$ & 1.21 \\[1.5mm]
\bottomrule
\end{tabular}
\end{center}
\tablefoot{
All masses are in units of $h^{-1} \mathrm{M}_\sun$ and correlation length $r_0$ is given in $h^{-1} \mathrm{Mpc}$.
}
\label{table:wp_params}
\end{table*}

We observe a marginal difference between clustering strength of the mergers and non-mergers, with the non-mergers showing a trend of stronger clustering than mergers. 
This is deducible from the higher value of the correlation length $r_0$ of the mergers in comparison with the non-mergers, although the shift in $r_0$ is within $1\sigma$ significance.
The mergers based on $f$ exhibit a correlation length of $r_0 = 6.11 \pm 0.55 \, h^{-1} \mathrm{Mpc}$, whereas the same for the corresponding non-mergers is $6.62 \pm 0.30 \, h^{-1} \mathrm{Mpc}$ which is a $0.8\sigma$ deviation.
In case of the $g$ classification, the mergers exhibit a correlation length of $r_0 = 6.43 \pm 1.07 \, h^{-1} \mathrm{Mpc}$ and the non-mergers show $r_0 = 6.44 \pm 0.29 \, h^{-1} \mathrm{Mpc}$.
The correlation lengths for mergers and non-mergers based on the $g$ parameter exhibit comparable values, indicating a lack of statistically significant difference in their clustering behaviour.

We see the similar effect in Fig.~\ref{fig:merger-nonmerger-wp-cross} also, where the galaxies that are equally classified as non-mergers by both $f$ and $g$ show stronger clustering than galaxies that are equally classified as mergers.
That is, the correlation length $r_0$ rises from $5.25 \pm 1.48 \, h^{-1} \mathrm{Mpc}$ for the sample $\mathcal{F}_\mathrm{M} \mathcal{G}_\mathrm{M}$ to $6.50 \pm 0.37 \, h^{-1} \mathrm{Mpc}$ for the sample $\mathcal{F}_\mathrm{NM} \mathcal{G}_\mathrm{NM}$.
This difference corresponds to $0.8\sigma$.

\subsection{Marked correlation functions}\label{sec:result_mcf}

\begin{figure}[t]
    \centering
    \includegraphics[width=\linewidth]{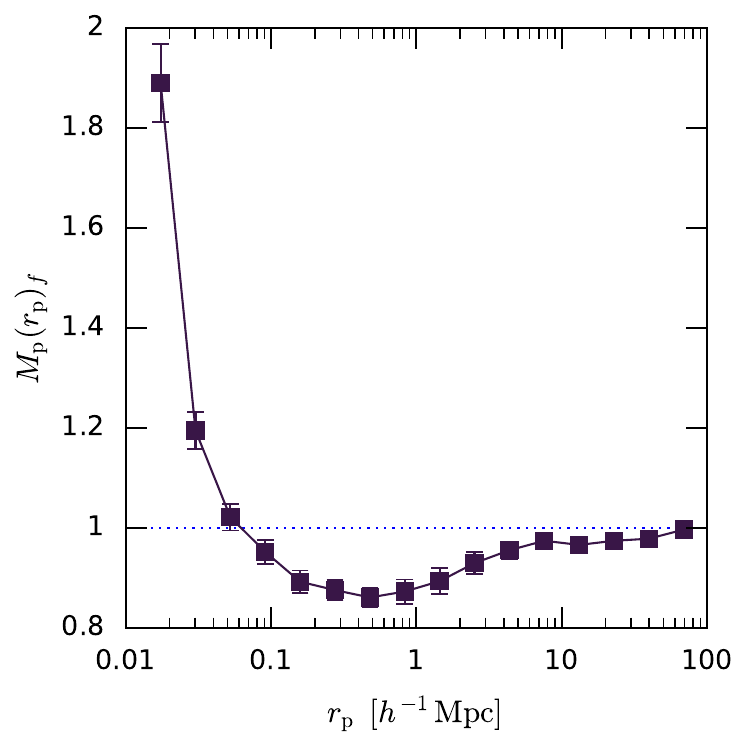}
    \caption{Marked correlation function measurements obtained using $f$ as marks.
    The error bar on each bin is a combination of its jackknife and random shuffling errors.}
    \label{fig:mcf_merger}
\end{figure}

In Fig.~\ref{fig:mcf_merger} we show the MCF computed using $f$ as marks in the main sample of 23855 galaxies from the GAMA survey.
Due to the superior number statistics of the main sample compared to the subsamples, more number of galaxy pairs contribute to the smaller separation bins. 
The smallest $r_\mathrm{p}$ bin for which we were able to obtain a reliable MCF measurement is centred at $\sim$ $0.01 h^{-1} \mathrm{Mpc}$.
We show the measurements for the projected separation scale of $0.01 < r_\mathrm{p}/h^{-1} \mathrm{Mpc} < 70$.
We combined the errors from the jackknife method and the random mark shuffling method and present them as error bars of the $M_\mathrm{p}(r_\mathrm{p})$ measurements, as described in Sect.~\ref{sec:measurement_mcf}.

We observe that the MCF with $f$ as marks strongly deviate from unity at all scales, which shows a strong correlation between that merger indicator and the environment.
We see an anti-correlation ($M_\mathrm{p} < 1$) in most bins and a strong correlation ($M_\mathrm{p} > 1$) in the separation scale $r_\mathrm{p} < 50 \, h^{-1} \mathrm{kpc}$.
This is discussed further in the next section.

\section{Discussion}\label{sec:discussion}

\subsection{Spatial clustering of galaxy mergers}\label{sec:discussion_envt_2pcf}

The 2pCF quantifies galaxy clustering and its dependence on galaxy properties. 
The correlation length $r_0$ reveals the strength of clustering: higher amplitude implies stronger clustering (see Sect.~\ref{sec:measurement_2pcf}). 
In Sect.~\ref{sec:result_2pcf} we compared the correlation lengths between galaxy mergers and non-mergers. 
We adopted two classification schemes to identify mergers: $f$ and $g$ (see Sect.~\ref{sec:data_mergindicators} for details). 
In the redshift range $0.1 < z < 0.15$, we do not observe a significant dependence of galaxy clustering on mergers. 
The correlation length $r_0$ obtained from the best-fitting HOD models do not differ significantly between mergers and non-mergers in case of both the classification schemes.

At the same time, we notice that galaxies classified as mergers are more massive and luminous than those classified as non-mergers.
For example, the sample $\mathcal{F}_\mathrm{M}$ contains galaxies with mean stellar mass $\log \left(M_{\star}/\mathrm{M}_{\sun} \right) = 10.27 \pm 0.45$ and mean $r$-band absolute magnitude $M_r = -20.95 \pm 0.90$ while $\mathcal{F}_\mathrm{NM}$ contains galaxies with mean stellar mass $\log \left(M_{\star}/\mathrm{M}_{\sun} \right) = 10.17 \pm 0.43$ and mean $r$-band absolute magnitude $M_r = -20.56 \pm 0.86$.
Similarly $\mathcal{G}_\mathrm{M}$ galaxies are more massive and more luminous than $\mathcal{G}_\mathrm{NM}$ galaxies.
It means that the selection of merging objects select the massive and brighter galaxies from the main sample.
Although these stellar mass differences are not statistically significant, the difference in correlation lengths between mergers and non-mergers could be partly driven by the difference in luminosity and stellar mass between the two populations. 

The splitting of the parent sample into mergers and non-mergers imposes additional property dependence of 2pCF.
This makes the study of environmental dependence of mergers using 2pCF complicated and hard to interpret.
To tackle this complexity, we choose to interpret the MCF measurements in which we do not split the parent sample.
By doing so, we minimise the influence of secondary dependencies and can utilise the parent sample with better number statistics.
This allows for more robust statistical analysis, enabling the investigation into the local environment of galaxy mergers at smaller separation scales.
This aspect is discussed in detail in Sects.~\ref{sec:discussion_envt_mcf} and \ref{sec:discussion_sfreffect}.

\subsection{Local environment of galaxy mergers}\label{sec:discussion_envt_mcf}

We used the MCF measurements with $f$ as the mark to study the local environment of galaxy mergers.
The $f$ value of a galaxy corresponds to the probability of that galaxy being a merger.
The choice of this mark is discussed in Sect.~\ref{sec:data_markchoice}.
The MCF at a given separation obtained using any property as the mark corresponds to the probability of finding galaxy pairs with that separation and both the galaxies in the pair being more pronounced in that property.
We emphasis that by `pairs', we do not refer to the pairs of galaxies that are actually merging with each other. 
Rather, we refer to the pairs (related or unrelated) that can be formed out of all the galaxies in the sample.

The anti-correlation observed in Fig.~\ref{fig:mcf_merger}, that is, $M_\mathrm{p}(r_\mathrm{p})_f < 1$ with an average significance of $3.4\sigma$ on scales $r_\mathrm{p} > 50 \, h^{-1} \mathrm{kpc}$ signifies a low probability of finding pairs of galaxies that are mergers and separated by scales greater than $50 \, h^{-1} \mathrm{kpc}$. 
This result confirms our finding from the 2pCF that galaxy mergers tend to exhibit less clustering than non-mergers. 
That is, our measurements imply that in the redshift range of $0.1 < z < 0.15$, galaxy mergers do not tend to exist in the high-density regions of the cosmic web.
This result can be explained based on the dependence of merging events on the relative velocities of galaxies.
Mergers tend to be more likely in regions with a higher concentration of galaxies with moderate relative velocities.
Too low velocities would take too long for the merging to happen, whereas too high velocities cause them to pass by each other, making it difficult for the galaxies to merge \citep{ostriker1980, mihos2004, jian2012, deger2018, benavides2020}.
Therefore, the reason for our observation that the galaxy mergers do not prefer the densest environments can be attributed to their high relative velocities that hinder the merging process.

\subsection{Halo properties of mergers}

The preference of galaxy mergers to occur in the less dense regions is also evident in the halo mass estimates from the HOD modelling of 2pCFs. 
It has been established that the densest environments are highly probable to be dominated by the most massive dark matter haloes, while the less dense regions by less massive haloes \citep[e.g.][]{mo1996, lemson1999}. 
As presented in Table~\ref{table:wp_params}, galaxy mergers prefer the latter. 
Based on the average halo masses estimated using best-fit HOD parameters, galaxy mergers are on average hosted by less massive dark matter haloes than non-mergers - for both merger samples $\log \langle M_\text{h} \rangle \sim 13.25$ $h^{-1} \mathrm{M}_\sun$, while for non-mergers $\log \langle M_\text{h} \rangle \sim 13.46$ $h^{-1} \mathrm{M}_\sun$. 
The difference is even more striking when we compare mergers selected by both the $f$ and $g$ parameters. 
For $\mathcal{F}_\mathrm{M} \mathcal{G}_\mathrm{M}$ the average halo mass hosting a mergers is $\log \langle M_\text{h} \rangle = 12.76$ $h^{-1} \mathrm{M}_\sun$ and for purely non-merger sample $\log \langle M_h \rangle = 13.50$ $h^{-1} \mathrm{M}_\sun$. 

Similarly the satellite fraction estimates support distinct environmental preferences of the mergers. 
However, the interpretation of these results is not straightforward and includes some caveats. 
Satellite fraction for merger samples is lower, at level of $\sim 0.31\%$ than for non mergers which is $\sim 53\%$. 
The difference is even bigger when we look at merger and non-mergers selected using both methods at the same time. 
Then satellite fraction for mergers is only $4\%$ and non-mergers is $34\%$ (see Table \ref{table:wp_params}). 
At a face value this might suggest that the mergers are hosted by less abundant, hence less densely populated haloes. 
However, lower value of $f_\text{s}$ for merging galaxies does not necessary mean that there are no other satellite galaxies occupying a dark matter halo, rather that there are no merging satellite galaxies within a given stellar mass limit imposed on a galaxy sample. 
Our samples cover wide range of stellar masses ($9.5 < \log \left(M_{\star}/\mathrm{M}_{\sun} \right) < \sim 12.0$) and the stellar mass distribution of merger and non-merger samples is similar (see mean stellar masses in Table~\ref{table:subsamples_properties}). 
Therefore low satellite fraction for merging galaxies suggest that it is more probable that haloes host non merging galaxies rather than mergers. 

The interpretation of our results within the HOD framework comes with some caveats. 
According to our measurements, galaxy clustering differs between mergers and non-mergers, consequently halo masses of these samples differ. 
This can be driven by the dependence of clustering on mergers, but also by the underlying correlation between mergers and other galaxy properties such as stellar mass and/or luminosity and/or SFR, which has been shown to influence clustering properties \citep[see e.g,][]{zheng2007_HOD, blanton&moustakas2009, zehavi2011}. 
There is no agreement in the literature about these dependencies. 
While an investigation into disentangling these dependencies is interesting and deserving of study, we believe it is beyond the scope of this paper. 
In our case the marked correlation function measurements and conclusions (see Sect.~\ref{sec:discussion_envt_mcf}) provide stronger arguments and are hence our preferred method to study merger-environment correlations.

\subsection{Comparison with previous works}

Our observation agrees with the conclusions of \citet{perez2009}, \citet{ellison2010}, and \citet{omori2023}.
\citet{perez2009} found that galaxy interactions and mergers are more frequent in intermediate-density environments.
\citet{ellison2010} observed that the high density environments are dominated by galaxy pairs with high relative velocities and hence less mergers.
\citet{omori2023} observed that galaxy mergers favour lower density environments, whereas non-mergers favour higher density environments at scales greater than 0.5 $h^{-1} \mathrm{Mpc}$.
It is to be noted that these works characterise the local environment using either the aperture method or the fifth nearest neighbour method, whereas we probe the environment simultaneously over a range of scales.
So, a quantitative comparison is not possible between our results and these works. 

We can compare our MCF measurements with that of \citet[][Fig.~4]{skibba2009_morphology}.
They measured MCF with $P_\mathrm{mg}$, the likelihood with which a galaxy is classified as undergoing a merger or an interaction, as mark.
They observed an insignificant MCF signal for scales $r_\mathrm{p} >  100 \, h^{-1} \mathrm{kpc}$, whereas we observe an anti-correlation for those scales.
Although their mark and our mark are equivalent in nature, the major differences between these two works are the redshift range and flux limits of the samples. 
Their sample covers a lower range ($0.017 < z < 0.082$) and has a brighter flux limit ($r < 17.8$) than ours.
We speculate that this disparity may be attributed to these differences between the samples.
Nevertheless, our measurements are consistent with them within the error bars.

\subsection{Enhancement of marked correlation function at smallest scales}\label{sec:discussion_sfreffect}

One important observation is the sudden rise in MCF with $f$ as mark for $r_\mathrm{p} < 50 \, h^{-1} \mathrm{kpc}$.
The $M_\mathrm{p}(r_\mathrm{p})$ signal for the smallest separation bin $r_\mathrm{p} = 17 \, h^{-1} \mathrm{kpc}$ reaches $1.89 \pm 0.06$, which means a significantly strong correlation with the environment.
We recall that the higher MCF with $f$ as mark at this separation signifies the higher probability of finding galaxy pairs with separation $\sim 17 \, h^{-1} \mathrm{kpc}$ and both the galaxies in the pairs having a higher value of $f$ than average.
It is to be noted that as we move to the smallest separation scales (of the order of $10 \, \mathrm{kpc}$), we are probing the typical scales at which galaxy-galaxy interactions happen. 
That means, there are high chances for galaxies in the pairs with separation $< 50 \, h^{-1} \mathrm{kpc}$ to be interacting or even merging with each other.
So both the galaxies in these close pairs carry a higher $f$ value and thereby contribute to the strong $f$-MCF on small scales. 
Such a peak for the smaller scales was observed by \citet{skibba2009_morphology} also in their $P_\mathrm{mg}$ MCF.

Interestingly, our MCF measurements with $f$ as mark are very similar to MCF measurements with SFR as mark by \citet{gunawardhana2018} and \citet{sureshkumar2023} which are based on the GAMA data.
Numerous studies observe enhanced SFR in merging galaxies due to gas compression \citep{li2008, robaina2009, darg2010_gz_prop, davies2015, pearson2019_merger_sfr, pearson2022, shah2022}.
So, the similarity in MCF measurements between SFR and merger probability suggests a similarity in the environmental correlations of the mergers and SFR.

\section{Conclusions}\label{sec:conclusions}

In this paper, we investigated whether galaxy mergers prefer to exist in the over-dense or under-dense regions of the LSS.
We used a set of galaxies in the redshift range $0.1 < z < 0.15$ from the GAMA survey.
We measured 2pCF of the mergers and non-mergers defined based on the CNN based classification parameter $f$ and the non-parametric morphological parameter $g$.
We investigated the environmental dependence of galaxy mergers using MCF measurements with galaxies weighted using $f$.

In our study using 2pCF, we did not observe a statistically significant difference between the clustering strengths of the mergers and non-mergers.
At the same time, we established a clear anti-correlation between the probability of a galaxy undergoing a merger and environment using MCF. 
Hence, the main conclusion of this work is that galaxy mergers prefer under-dense environments of the cosmic web in the local Universe. 
We attribute this observation to the high relative velocities of galaxy pairs in densely populated environments, which restrict the likelihood of mergers. 

Additionally, based on the comparison between MCF signals with merger probability as mark and our previous MCF signals with SFR as mark \citep{sureshkumar2023}, we suggested the similar environmental correlations of galaxy mergers and SFR.

In summary, the galaxy mergers appear to be correlated with the local environment of the LSS.
We demonstrated the advantage of MCF over 2pCF to explore these environmental correlations.
This study complements previous studies on the correlation between galaxy mergers and the local environment.
Our results contribute towards further investigations on the role of mergers in galaxy evolution. 
We plan to extend our measurements using simulations and high-redshift surveys, with which we hope to better understand the evolution of galaxy mergers over cosmic time and in different environments.

\begin{acknowledgements}
We thank the anonymous referee for the useful comments and suggestions.
We thank Matt Hilton for his comments on the draft.
U.S. acknowledges support from the National Research Foundation of South Africa (grant no. 137975).
W.J.P. has been supported by the Polish National Science Center project UMO-2020/37/B/ST9/00466 and by the Foundation for Polish Science (FNP).
This research was supported by the Polish National Science Centre grant UMO-2018/30/M/ST9/00757 and by Polish Ministry of Science and Higher Education grant DIR/WK/2018/12.
GAMA is a joint European-Australasian project based around a spectroscopic campaign using the Anglo-Australian Telescope. The GAMA input catalogue is based on data taken from the Sloan Digital Sky Survey and the UKIRT Infrared Deep Sky Survey. Complementary imaging of the GAMA regions is being obtained by a number of independent survey programmes including GALEX MIS, VST KiDS, VISTA VIKING, WISE, Herschel-ATLAS, GMRT, and ASKAP providing UV to radio coverage. GAMA is funded by the STFC (UK), the ARC (Australia), the AAO, and the participating institutions. The GAMA website is \url{http://www.gama-survey.org/}.
During this research, we made use of Tool for OPerations on Catalogues And Tables \citep[TOPCAT;][]{taylor2005_topcat} and NASA’s Astrophysics Data System Bibliographic Services.
This research made use of \textsc{HaloMod}\footnote{https://pypi.org/project/halomod/} \citep{murray2021} and \textsc{Emcee} \footnote{https://emcee.readthedocs.io/en/stable/} \citep{foreman2013}.
We gratefully acknowledge Polish high-performance computing infrastructure PLGrid (HPC Center: ACK Cyfronet AGH) for providing computer facilities and support within computational grant no. PLG/2023/016554.
We also acknowledge the use of OAUJ cluster computing facility and Hippo at the University of KwaZulu-Natal.
\end{acknowledgements}

\bibliographystyle{aa} 
\bibliography{references} 

\begin{thebibliography}{123}
\expandafter\ifx\csname natexlab\endcsname\relax\def\natexlab#1{#1}\fi

\bibitem[{{Abraham} {et~al.}(1996){Abraham}, {Tanvir}, {Santiago}, {Ellis},
  {Glazebrook}, \& {van den Bergh}}]{abraham1996}
{Abraham}, R.~G., {Tanvir}, N.~R., {Santiago}, B.~X., {et~al.} 1996, \mnras,
  279, L47

\bibitem[{{Alonso} {et~al.}(2012){Alonso}, {Mesa}, {Padilla}, \&
  {Lambas}}]{alonso2012}
{Alonso}, S., {Mesa}, V., {Padilla}, N., \& {Lambas}, D.~G. 2012, \aap, 539,
  A46

\bibitem[{{Baldry} {et~al.}(2010){Baldry}, {Robotham}, {Hill}, {Driver},
  {Liske}, {Norberg}, {Bamford}, {Hopkins}, {Loveday}, {Peacock}, {Cameron},
  {Croom}, {Cross}, {Doyle}, {Dye}, {Frenk}, {Jones}, {van Kampen}, {Kelvin},
  {Nichol}, {Parkinson}, {Popescu}, {Prescott}, {Sharp}, {Sutherland},
  {Thomas}, \& {Tuffs}}]{baldry2010}
{Baldry}, I.~K., {Robotham}, A.~S.~G., {Hill}, D.~T., {et~al.} 2010, \mnras,
  404, 86

\bibitem[{{Barnes}(1992)}]{barnes1992}
{Barnes}, J.~E. 1992, \apj, 393, 484

\bibitem[{{Barnes} \& {Hernquist}(1991)}]{barnes1991}
{Barnes}, J.~E. \& {Hernquist}, L.~E. 1991, \apjl, 370, L65

\bibitem[{{Beisbart} \& {Kerscher}(2000)}]{beisbart2000}
{Beisbart}, C. \& {Kerscher}, M. 2000, \apj, 545, 6

\bibitem[{{Benavides} {et~al.}(2020){Benavides}, {Sales}, \&
  {Abadi}}]{benavides2020}
{Benavides}, J.~A., {Sales}, L.~V., \& {Abadi}, M.~G. 2020, \mnras, 498, 3852

\bibitem[{{Blanton} \& {Moustakas}(2009)}]{blanton&moustakas2009}
{Blanton}, M.~R. \& {Moustakas}, J. 2009, \araa, 47, 159

\bibitem[{{Bruzual} \& {Charlot}(2003)}]{bruzual2003}
{Bruzual}, G. \& {Charlot}, S. 2003, \mnras, 344, 1000

\bibitem[{{Calzetti} {et~al.}(2000){Calzetti}, {Armus}, {Bohlin}, {Kinney},
  {Koornneef}, \& {Storchi-Bergmann}}]{calzetti2000}
{Calzetti}, D., {Armus}, L., {Bohlin}, R.~C., {et~al.} 2000, \apj, 533, 682

\bibitem[{{Chabrier}(2003)}]{chabrier2003}
{Chabrier}, G. 2003, \apjl, 586, L133

\bibitem[{{Clontz} {et~al.}(2022){Clontz}, {Wake}, \& {Zheng}}]{clontz2022}
{Clontz}, C., {Wake}, D., \& {Zheng}, Z. 2022, \mnras, 515, 2224

\bibitem[{{Coil} {et~al.}(2008){Coil}, {Newman}, {Croton}, {Cooper}, {Davis},
  {Faber}, {Gerke}, {Koo}, {Padmanabhan}, {Wechsler}, \& {Weiner}}]{coil2008}
{Coil}, A.~L., {Newman}, J.~A., {Croton}, D., {et~al.} 2008, \apj, 672, 153

\bibitem[{{Cole}(2011)}]{cole2011}
{Cole}, S. 2011, \mnras, 416, 739

\bibitem[{{Conselice} {et~al.}(2003){Conselice}, {Bershady}, {Dickinson}, \&
  {Papovich}}]{conselice2003}
{Conselice}, C.~J., {Bershady}, M.~A., {Dickinson}, M., \& {Papovich}, C. 2003,
  \aj, 126, 1183

\bibitem[{{Conselice} {et~al.}(2000){Conselice}, {Bershady}, \&
  {Jangren}}]{conselice2000}
{Conselice}, C.~J., {Bershady}, M.~A., \& {Jangren}, A. 2000, \apj, 529, 886

\bibitem[{{Cooray} \& {Sheth}(2002)}]{cooray2002}
{Cooray}, A. \& {Sheth}, R. 2002, \physrep, 372, 1

\bibitem[{{Coupon} {et~al.}(2012){Coupon}, {Kilbinger}, {McCracken}, {Ilbert},
  {Arnouts}, {Mellier}, {Abbas}, {de la Torre}, {Goranova}, {Hudelot}, {Kneib},
  \& {Le F{\`e}vre}}]{coupon2012}
{Coupon}, J., {Kilbinger}, M., {McCracken}, H.~J., {et~al.} 2012, \aap, 542, A5

\bibitem[{{Darg} {et~al.}(2010{\natexlab{a}}){Darg}, {Kaviraj}, {Lintott},
  {Schawinski}, {Sarzi}, {Bamford}, {Silk}, {Andreescu}, {Murray}, {Nichol},
  {Raddick}, {Slosar}, {Szalay}, {Thomas}, \& {Vandenberg}}]{darg2010_gz_prop}
{Darg}, D.~W., {Kaviraj}, S., {Lintott}, C.~J., {et~al.} 2010{\natexlab{a}},
  \mnras, 401, 1552

\bibitem[{{Darg} {et~al.}(2010{\natexlab{b}}){Darg}, {Kaviraj}, {Lintott},
  {Schawinski}, {Sarzi}, {Bamford}, {Silk}, {Proctor}, {Andreescu}, {Murray},
  {Nichol}, {Raddick}, {Slosar}, {Szalay}, {Thomas}, \&
  {Vandenberg}}]{darg2010_gz_sdss}
{Darg}, D.~W., {Kaviraj}, S., {Lintott}, C.~J., {et~al.} 2010{\natexlab{b}},
  \mnras, 401, 1043

\bibitem[{{Davies} {et~al.}(2015){Davies}, {Robotham}, {Driver}, {Alpaslan},
  {Baldry}, {Bland-Hawthorn}, {Brough}, {Brown}, {Cluver}, {Drinkwater},
  {Foster}, {Grootes}, {Konstantopoulos}, {Lara-L{\'o}pez},
  {L{\'o}pez-S{\'a}nchez}, {Loveday}, {Meyer}, {Moffett}, {Norberg}, {Owers},
  {Popescu}, {De Propris}, {Sharp}, {Tuffs}, {Wang}, {Wilkins}, {Dunne},
  {Bourne}, \& {Smith}}]{davies2015}
{Davies}, L.~J.~M., {Robotham}, A.~S.~G., {Driver}, S.~P., {et~al.} 2015,
  \mnras, 452, 616

\bibitem[{{de Jong} {et~al.}(2013){de Jong}, {Verdoes Kleijn}, {Kuijken}, \&
  {Valentijn}}]{dejong2013_kids}
{de Jong}, J. T.~A., {Verdoes Kleijn}, G.~A., {Kuijken}, K.~H., \& {Valentijn},
  E.~A. 2013, Experimental Astronomy, 35, 25

\bibitem[{{de la Torre} {et~al.}(2007){de la Torre}, {Le F{\`e}vre}, {Arnouts},
  {Guzzo}, {Farrah}, {Iovino}, {Lonsdale}, {Meneux}, {Oliver}, {Pollo},
  {Waddington}, {Bottini}, {Fang}, {Garilli}, {Le Brun}, {Maccagni}, {Picat},
  {Scaramella}, {Scodeggio}, {Shupe}, {Surace}, {Tresse}, {Vettolani},
  {Zanichelli}, {Adami}, {Bardelli}, {Bolzonella}, {Cappi}, {Charlot},
  {Ciliegi}, {Contini}, {Foucaud}, {Franzetti}, {Gavignaud}, {Ilbert},
  {Lamareille}, {McCracken}, {Marano}, {Marinoni}, {Mazure}, {Merighi},
  {Paltani}, {Pell{\`o}}, {Pozzetti}, {Radovich}, {Zamorani}, {Zucca}, {Bondi},
  {Bongiorno}, {Brinchmann}, {Cucciati}, {Mellier}, {Merluzzi}, {Temporin},
  {Vergani}, \& {Walcher}}]{torre2007}
{de la Torre}, S., {Le F{\`e}vre}, O., {Arnouts}, S., {et~al.} 2007, \aap, 475,
  443

\bibitem[{{de Ravel} {et~al.}(2011){de Ravel}, {Kampczyk}, {Le F{\`e}vre},
  {Lilly}, {Tasca}, {Tresse}, {Lopez-Sanjuan}, {Bolzonella}, {Kovac}, {Abbas},
  {Bardelli}, {Bongiorno}, {Caputi}, {Contini}, {Coppa}, {Cucciati}, {de la
  Torre}, {Dunlop}, {Franzetti}, {Garilli}, {Iovino}, {Kneib}, {Koekemoer},
  {Knobel}, {Lamareille}, {Le Borgne}, {Le Brun}, {Leauthaud}, {Maier},
  {Mainieri}, {Mignoli}, {Pello}, {Peng}, {Perez Montero}, {Ricciardelli},
  {Scodeggio}, {Silverman}, {Tanaka}, {Vergani}, {Zamorani}, {Zucca},
  {Bottini}, {Cappi}, {Carollo}, {Cassata}, {Cimatti}, {Fumana}, {Guzzo},
  {Maccagni}, {Marinoni}, {McCracken}, {Memeo}, {Meneux}, {Oesch}, {Porciani},
  {Pozzetti}, {Renzini}, {Scaramella}, \& {Scarlata}}]{deravel2011}
{de Ravel}, L., {Kampczyk}, P., {Le F{\`e}vre}, O., {et~al.} 2011, arXiv
  e-prints, arXiv:1104.5470

\bibitem[{{Deger} {et~al.}(2018){Deger}, {Rudnick}, {Kelkar},
  {Arag{\'o}n-Salamanca}, {Desai}, {Lotz}, {Jablonka}, {Moustakas}, \&
  {Zaritsky}}]{deger2018}
{Deger}, S., {Rudnick}, G., {Kelkar}, K., {et~al.} 2018, \apj, 869, 6

\bibitem[{{Delahaye} {et~al.}(2017){Delahaye}, {Webb}, {Nantais}, {DeGroot},
  {Wilson}, {Muzzin}, {Yee}, {Foltz}, {Noble}, {Demarco}, {Tudorica}, {Cooper},
  {Lidman}, {Perlmutter}, {Hayden}, {Boone}, \& {Surace}}]{delahaye2017}
{Delahaye}, A.~G., {Webb}, T.~M.~A., {Nantais}, J., {et~al.} 2017, \apj, 843,
  126

\bibitem[{{Dotti} {et~al.}(2012){Dotti}, {Sesana}, \& {Decarli}}]{dotti2012}
{Dotti}, M., {Sesana}, A., \& {Decarli}, R. 2012, Advances in Astronomy, 2012,
  940568

\bibitem[{{Driver} {et~al.}(2022){Driver}, {Bellstedt}, {Robotham}, {Baldry},
  {Davies}, {Liske}, {Obreschkow}, {Taylor}, {Wright}, {Alpaslan}, {Bamford},
  {Bauer}, {Bland-Hawthorn}, {Bilicki}, {Bravo}, {Brough}, {Casura}, {Cluver},
  {Colless}, {Conselice}, {Croom}, {de Jong}, {D'Eugenio}, {De Propris},
  {Dogruel}, {Drinkwater}, {Dvornik}, {Farrow}, {Frenk}, {Giblin}, {Graham},
  {Grootes}, {Gunawardhana}, {Hashemizadeh}, {H{\"a}u{\ss}ler}, {Heymans},
  {Hildebrandt}, {Holwerda}, {Hopkins}, {Jarrett}, {Heath Jones}, {Kelvin},
  {Koushan}, {Kuijken}, {Lara-L{\'o}pez}, {Lange}, {L{\'o}pez-S{\'a}nchez},
  {Loveday}, {Mahajan}, {Meyer}, {Moffett}, {Napolitano}, {Norberg}, {Owers},
  {Radovich}, {Raouf}, {Peacock}, {Phillipps}, {Pimbblet}, {Popescu}, {Said},
  {Sansom}, {Seibert}, {Sutherland}, {Thorne}, {Tuffs}, {Turner}, {van der
  Wel}, {van Kampen}, \& {Wilkins}}]{driver2022_gamadr4}
{Driver}, S.~P., {Bellstedt}, S., {Robotham}, A. S.~G., {et~al.} 2022, \mnras,
  513, 439

\bibitem[{{Driver} {et~al.}(2011){Driver}, {Hill}, {Kelvin}, {Robotham},
  {Liske}, {Norberg}, {Baldry}, {Bamford}, {Hopkins}, {Loveday}, {Peacock},
  {Andrae}, {Bland-Hawthorn}, {Brough}, {Brown}, {Cameron}, {Ching}, {Colless},
  {Conselice}, {Croom}, {Cross}, {de Propris}, {Dye}, {Drinkwater}, {Ellis},
  {Graham}, {Grootes}, {Gunawardhana}, {Jones}, {van Kampen}, {Maraston},
  {Nichol}, {Parkinson}, {Phillipps}, {Pimbblet}, {Popescu}, {Prescott},
  {Roseboom}, {Sadler}, {Sansom}, {Sharp}, {Smith}, {Taylor}, {Thomas},
  {Tuffs}, {Wijesinghe}, {Dunne}, {Frenk}, {Jarvis}, {Madore}, {Meyer},
  {Seibert}, {Staveley-Smith}, {Sutherland}, \&
  {Warren}}]{driver2011_gama_coredata}
{Driver}, S.~P., {Hill}, D.~T., {Kelvin}, L.~S., {et~al.} 2011, \mnras, 413,
  971

\bibitem[{{Driver} {et~al.}(2009){Driver}, {Norberg}, {Baldry}, {Bamford},
  {Hopkins}, {Liske}, {Loveday}, {Peacock}, {Hill}, {Kelvin}, {Robotham},
  {Cross}, {Parkinson}, {Prescott}, {Conselice}, {Dunne}, {Brough}, {Jones},
  {Sharp}, {van Kampen}, {Oliver}, {Roseboom}, {Bland-Hawthorn}, {Croom},
  {Ellis}, {Cameron}, {Cole}, {Frenk}, {Couch}, {Graham}, {Proctor}, {De
  Propris}, {Doyle}, {Edmondson}, {Nichol}, {Thomas}, {Eales}, {Jarvis},
  {Kuijken}, {Lahav}, {Madore}, {Seibert}, {Meyer}, {Staveley-Smith},
  {Phillipps}, {Popescu}, {Sansom}, {Sutherland}, {Tuffs}, \&
  {Warren}}]{driver2009_gama_gen}
{Driver}, S.~P., {Norberg}, P., {Baldry}, I.~K., {et~al.} 2009, Astronomy and
  Geophysics, 50, 5.12

\bibitem[{{Duncan} {et~al.}(2019){Duncan}, {Conselice}, {Mundy}, {Bell},
  {Donley}, {Galametz}, {Guo}, {Grogin}, {Hathi}, {Kartaltepe}, {Kocevski},
  {Koekemoer}, {P{\'e}rez-Gonz{\'a}lez}, {Mantha}, {Snyder}, \&
  {Stefanon}}]{duncan2019}
{Duncan}, K., {Conselice}, C.~J., {Mundy}, C., {et~al.} 2019, \apj, 876, 110

\bibitem[{{Durkalec} {et~al.}(2018){Durkalec}, {Le F{\`e}vre}, {Pollo},
  {Zamorani}, {Lemaux}, {Garilli}, {Bardelli}, {Hathi}, {Koekemoer}, {Pforr},
  \& {Zucca}}]{durkalec2018}
{Durkalec}, A., {Le F{\`e}vre}, O., {Pollo}, A., {et~al.} 2018, \aap, 612, A42

\bibitem[{{Ellison} {et~al.}(2008){Ellison}, {Patton}, {Simard}, \&
  {McConnachie}}]{ellison2008}
{Ellison}, S.~L., {Patton}, D.~R., {Simard}, L., \& {McConnachie}, A.~W. 2008,
  \aj, 135, 1877

\bibitem[{{Ellison} {et~al.}(2010){Ellison}, {Patton}, {Simard}, {McConnachie},
  {Baldry}, \& {Mendel}}]{ellison2010}
{Ellison}, S.~L., {Patton}, D.~R., {Simard}, L., {et~al.} 2010, \mnras, 407,
  1514

\bibitem[{{Ellison} {et~al.}(2019){Ellison}, {Viswanathan}, {Patton},
  {Bottrell}, {McConnachie}, {Gwyn}, \& {Cuillandre}}]{ellison2019}
{Ellison}, S.~L., {Viswanathan}, A., {Patton}, D.~R., {et~al.} 2019, \mnras,
  487, 2491

\bibitem[{{Fakhouri} \& {Ma}(2009)}]{fakhouri2009}
{Fakhouri}, O. \& {Ma}, C.-P. 2009, \mnras, 394, 1825

\bibitem[{{Farrow} {et~al.}(2015){Farrow}, {Cole}, {Norberg}, {Metcalfe},
  {Baldry}, {Bland-Hawthorn}, {Brown}, {Hopkins}, {Lacey}, {Liske}, {Loveday},
  {Palamara}, {Robotham}, \& {Sridhar}}]{farrow2015_gama_cf}
{Farrow}, D.~J., {Cole}, S., {Norberg}, P., {et~al.} 2015, \mnras, 454, 2120

\bibitem[{{Ferreira} {et~al.}(2020){Ferreira}, {Conselice}, {Duncan}, {Cheng},
  {Griffiths}, \& {Whitney}}]{ferreira2020}
{Ferreira}, L., {Conselice}, C.~J., {Duncan}, K., {et~al.} 2020, \apj, 895, 115

\bibitem[{{Foreman-Mackey} {et~al.}(2013){Foreman-Mackey}, {Hogg}, {Lang}, \&
  {Goodman}}]{foreman2013}
{Foreman-Mackey}, D., {Hogg}, D.~W., {Lang}, D., \& {Goodman}, J. 2013, \pasp,
  125, 306

\bibitem[{{Gunawardhana} {et~al.}(2018){Gunawardhana}, {Norberg}, {Zehavi},
  {Farrow}, {Loveday}, {Hopkins}, {Davies}, {Wang}, {Alpaslan}, {Bland-
  Hawthorn}, {Brough}, {Holwerda}, {Owers}, \& {Wright}}]{gunawardhana2018}
{Gunawardhana}, M.~L.~P., {Norberg}, P., {Zehavi}, I., {et~al.} 2018, \mnras,
  479, 1433

\bibitem[{{Hartlap} {et~al.}(2007){Hartlap}, {Simon}, \&
  {Schneider}}]{hartlap2007}
{Hartlap}, J., {Simon}, P., \& {Schneider}, P. 2007, \aap, 464, 399

\bibitem[{{Hartley} {et~al.}(2010){Hartley}, {Almaini}, {Cirasuolo}, {Foucaud},
  {Simpson}, {Conselice}, {Smail}, {McLure}, {Dunlop}, {Chuter}, {Maddox},
  {Lane}, \& {Bradshaw}}]{hartley2010}
{Hartley}, W.~G., {Almaini}, O., {Cirasuolo}, M., {et~al.} 2010, \mnras, 407,
  1212

\bibitem[{{Heinis} {et~al.}(2004){Heinis}, {Treyer}, {Arnouts}, {Milliard},
  {Donas}, {Gal}, {Martin}, \& {Viton}}]{heinis2004}
{Heinis}, S., {Treyer}, M., {Arnouts}, S., {et~al.} 2004, \aap, 424, L9

\bibitem[{{Hester} \& {Tasitsiomi}(2010)}]{hester2010}
{Hester}, J.~A. \& {Tasitsiomi}, A. 2010, \apj, 715, 342

\bibitem[{{Infante}(1994)}]{infante1994}
{Infante}, L. 1994, \aap, 282, 353

\bibitem[{{Jian} {et~al.}(2012){Jian}, {Lin}, \& {Chiueh}}]{jian2012}
{Jian}, H.-Y., {Lin}, L., \& {Chiueh}, T. 2012, \apj, 754, 26

\bibitem[{{Kaiser}(1987)}]{kaiser1987}
{Kaiser}, N. 1987, \mnras, 227, 1

\bibitem[{{Kampczyk} {et~al.}(2013){Kampczyk}, {Lilly}, {de Ravel}, {Le
  F{\`e}vre}, {Bolzonella}, {Carollo}, {Diener}, {Knobel}, {Kova{\v{c}}},
  {Maier}, {Renzini}, {Sargent}, {Vergani}, {Abbas}, {Bardelli}, {Bongiorno},
  {Bordoloi}, {Caputi}, {Contini}, {Coppa}, {Cucciati}, {de la Torre},
  {Franzetti}, {Garilli}, {Iovino}, {Kneib}, {Koekemoer}, {Lamareille}, {Le
  Borgne}, {Le Brun}, {Leauthaud}, {Mainieri}, {Mignoli}, {Pello}, {Peng},
  {Perez Montero}, {Ricciardelli}, {Scodeggio}, {Silverman}, {Tanaka}, {Tasca},
  {Tresse}, {Zamorani}, {Zucca}, {Bottini}, {Cappi}, {Cassata}, {Cimatti},
  {Fumana}, {Guzzo}, {Kartaltepe}, {Marinoni}, {McCracken}, {Memeo}, {Meneux},
  {Oesch}, {Porciani}, {Pozzetti}, \& {Scaramella}}]{kampczyk2013}
{Kampczyk}, P., {Lilly}, S.~J., {de Ravel}, L., {et~al.} 2013, \apj, 762, 43

\bibitem[{{Landy} \& {Szalay}(1993)}]{landy&szalay1993}
{Landy}, S.~D. \& {Szalay}, A.~S. 1993, \apj, 412, 64

\bibitem[{{Larson} \& {Tinsley}(1978)}]{larson1978}
{Larson}, R.~B. \& {Tinsley}, B.~M. 1978, \apj, 219, 46

\bibitem[{{Le F{\`e}vre} {et~al.}(2000){Le F{\`e}vre}, {Abraham}, {Lilly},
  {Ellis}, {Brinchmann}, {Schade}, {Tresse}, {Colless}, {Crampton},
  {Glazebrook}, {Hammer}, \& {Broadhurst}}]{lefevre2000}
{Le F{\`e}vre}, O., {Abraham}, R., {Lilly}, S.~J., {et~al.} 2000, \mnras, 311,
  565

\bibitem[{{LeCun} {et~al.}(2015){LeCun}, {Bengio}, \& {Hinton}}]{lecun2015}
{LeCun}, Y., {Bengio}, Y., \& {Hinton}, G. 2015, \nat, 521, 436

\bibitem[{{Lemson} \& {Kauffmann}(1999)}]{lemson1999}
{Lemson}, G. \& {Kauffmann}, G. 1999, \mnras, 302, 111

\bibitem[{{Li} {et~al.}(2008){Li}, {Kauffmann}, {Heckman}, {Jing}, \&
  {White}}]{li2008}
{Li}, C., {Kauffmann}, G., {Heckman}, T.~M., {Jing}, Y.~P., \& {White}, S.
  D.~M. 2008, \mnras, 385, 1903

\bibitem[{{Lin} {et~al.}(2010){Lin}, {Cooper}, {Jian}, {Koo}, {Patton}, {Yan},
  {Willmer}, {Coil}, {Chiueh}, {Croton}, {Gerke}, {Lotz}, {Guhathakurta}, \&
  {Newman}}]{lin2010}
{Lin}, L., {Cooper}, M.~C., {Jian}, H.-Y., {et~al.} 2010, \apj, 718, 1158

\bibitem[{{Lin} {et~al.}(2012){Lin}, {Dickinson}, {Jian}, {Merson}, {Baugh},
  {Scott}, {Foucaud}, {Wang}, {Yan}, {Yan}, {Cheng}, {Guo}, {Helly}, {Kirsten},
  {Koo}, {Lagos}, {Meger}, {Messias}, {Pope}, {Simard}, {Grogin}, \&
  {Wang}}]{lin2012}
{Lin}, L., {Dickinson}, M., {Jian}, H.-Y., {et~al.} 2012, \apj, 756, 71

\bibitem[{{Lintott} {et~al.}(2008){Lintott}, {Schawinski}, {Slosar}, {Land},
  {Bamford}, {Thomas}, {Raddick}, {Nichol}, {Szalay}, {Andreescu}, {Murray}, \&
  {Vandenberg}}]{lintott2008_galaxy_zoo}
{Lintott}, C.~J., {Schawinski}, K., {Slosar}, A., {et~al.} 2008, \mnras, 389,
  1179

\bibitem[{{Liske} {et~al.}(2015){Liske}, {Baldry}, {Driver}, {Tuffs},
  {Alpaslan}, {Andrae}, {Brough}, {Cluver}, {Grootes}, {Gunawardhana},
  {Kelvin}, {Loveday}, {Robotham}, {Taylor}, {Bamford}, {Bland- Hawthorn},
  {Brown}, {Drinkwater}, {Hopkins}, {Meyer}, {Norberg}, {Peacock}, {Agius},
  {Andrews}, {Bauer}, {Ching}, {Colless}, {Conselice}, {Croom}, {Davies}, {De
  Propris}, {Dunne}, {Eardley}, {Ellis}, {Foster}, {Frenk}, {H{\"a}u{\ss}ler},
  {Holwerda}, {Howlett}, {Ibarra}, {Jarvis}, {Jones}, {Kafle}, {Lacey},
  {Lange}, {Lara-L{\'o}pez}, {L{\'o}pez-S{\'a}nchez}, {Maddox}, {Madore},
  {McNaught-Roberts}, {Moffett}, {Nichol}, {Owers}, {Palamara}, {Penny},
  {Phillipps}, {Pimbblet}, {Popescu}, {Prescott}, {Proctor}, {Sadler},
  {Sansom}, {Seibert}, {Sharp}, {Sutherland}, {V{\'a }zquez-Mata}, {van
  Kampen}, {Wilkins}, {Williams}, \& {Wright}}]{liske2015_gama_dr2}
{Liske}, J., {Baldry}, I.~K., {Driver}, S.~P., {et~al.} 2015, \mnras, 452, 2087

\bibitem[{{Lotz} {et~al.}(2008{\natexlab{a}}){Lotz}, {Davis}, {Faber},
  {Guhathakurta}, {Gwyn}, {Huang}, {Koo}, {Le Floc'h}, {Lin}, {Newman},
  {Noeske}, {Papovich}, {Willmer}, {Coil}, {Conselice}, {Cooper}, {Hopkins},
  {Metevier}, {Primack}, {Rieke}, \& {Weiner}}]{lotz2008_gm20_def}
{Lotz}, J.~M., {Davis}, M., {Faber}, S.~M., {et~al.} 2008{\natexlab{a}}, \apj,
  672, 177

\bibitem[{{Lotz} {et~al.}(2008{\natexlab{b}}){Lotz}, {Jonsson}, {Cox}, \&
  {Primack}}]{lotz2008_merger_timescale}
{Lotz}, J.~M., {Jonsson}, P., {Cox}, T.~J., \& {Primack}, J.~R.
  2008{\natexlab{b}}, \mnras, 391, 1137

\bibitem[{{Lotz} {et~al.}(2004){Lotz}, {Primack}, \& {Madau}}]{lotz2004}
{Lotz}, J.~M., {Primack}, J., \& {Madau}, P. 2004, \aj, 128, 163

\bibitem[{{Mandelbaum} {et~al.}(2006){Mandelbaum}, {Hirata}, {Ishak}, {Seljak},
  \& {Brinkmann}}]{mandelbaum2006_JKNr}
{Mandelbaum}, R., {Hirata}, C.~M., {Ishak}, M., {Seljak}, U., \& {Brinkmann},
  J. 2006, \mnras, 367, 611

\bibitem[{{McCracken} {et~al.}(2015){McCracken}, {Wolk}, {Colombi},
  {Kilbinger}, {Ilbert}, {Peirani}, {Coupon}, {Dunlop}, {Milvang-Jensen},
  {Caputi}, {Aussel}, {B{\'e}thermin}, \& {Le F{\`e}vre}}]{mccracken2015}
{McCracken}, H.~J., {Wolk}, M., {Colombi}, S., {et~al.} 2015, \mnras, 449, 901

\bibitem[{{McIntosh} {et~al.}(2008){McIntosh}, {Guo}, {Hertzberg}, {Katz},
  {Mo}, {van den Bosch}, \& {Yang}}]{mcintosh2008}
{McIntosh}, D.~H., {Guo}, Y., {Hertzberg}, J., {et~al.} 2008, \mnras, 388, 1537

\bibitem[{{Meneux} {et~al.}(2006){Meneux}, {Le F{\`e}vre}, {Guzzo}, {Pollo},
  {Cappi}, {Ilbert}, {Iovino}, {Marinoni}, {McCracken}, {Bottini}, {Garilli},
  {Le Brun}, {Maccagni}, {Picat}, {Scaramella}, {Scodeggio}, {Tresse},
  {Vettolani}, {Zanichelli}, {Adami}, {Arnouts}, {Arnaboldi}, {Bardelli},
  {Bolzonella}, {Charlot}, {Ciliegi}, {Contini}, {Foucaud}, {Franzetti},
  {Gavignaud}, {Marano}, {Mazure}, {Merighi}, {Paltani}, {Pell{\`o}},
  {Pozzetti}, {Radovich}, {Zamorani}, {Zucca}, {Bondi}, {Bongiorno},
  {Busarello}, {Cucciati}, {Gregorini}, {Lamareille}, {Mathez}, {Mellier},
  {Merluzzi}, {Ripepi}, \& {Rizzo}}]{meneux2006}
{Meneux}, B., {Le F{\`e}vre}, O., {Guzzo}, L., {et~al.} 2006, \aap, 452, 387

\bibitem[{{Mihos}(2004)}]{mihos2004}
{Mihos}, J.~C. 2004, in Recycling Intergalactic and Interstellar Matter, Vol.
  217, Recycling Intergalactic and Interstellar Matter, ed. P.-A. {Duc},
  J.~{Braine}, \& E.~{Brinks}, 390

\bibitem[{{Mihos} \& {Hernquist}(1996)}]{mihos1996}
{Mihos}, J.~C. \& {Hernquist}, L. 1996, \apj, 464, 641

\bibitem[{{Milliard} {et~al.}(2007){Milliard}, {Heinis}, {Blaizot}, {Arnouts},
  {Schiminovich}, {Budav{\'a}ri}, {Donas}, {Treyer}, {Laget}, {Viton}, {Wyder},
  {Szalay}, {Barlow}, {Forster}, {Friedman}, {Martin}, {Morrissey}, {Neff},
  {Seibert}, {Small}, {Bianchi}, {Heckman}, {Lee}, {Madore}, {Rich}, {Welsh},
  {Yi}, \& {Xu}}]{milliard2007}
{Milliard}, B., {Heinis}, S., {Blaizot}, J., {et~al.} 2007, \apjs, 173, 494

\bibitem[{{Mo} \& {White}(1996)}]{mo1996}
{Mo}, H.~J. \& {White}, S.~D.~M. 1996, \mnras, 282, 347

\bibitem[{{Mostek} {et~al.}(2013){Mostek}, {Coil}, {Cooper}, {Davis}, {Newman},
  \& {Weiner}}]{mostek2013}
{Mostek}, N., {Coil}, A.~L., {Cooper}, M., {et~al.} 2013, \apj, 767, 89

\bibitem[{{Muldrew} {et~al.}(2012){Muldrew}, {Croton}, {Skibba}, {Pearce},
  {Ann}, {Baldry}, {Brough}, {Choi}, {Conselice}, {Cowan}, {Gallazzi}, {Gray},
  {Gr{\"u}tzbauch}, {Li}, {Park}, {Pilipenko}, {Podgorzec}, {Robotham},
  {Wilman}, {Yang}, {Zhang}, \& {Zibetti}}]{muldrew2012}
{Muldrew}, S.~I., {Croton}, D.~J., {Skibba}, R.~A., {et~al.} 2012, \mnras, 419,
  2670

\bibitem[{{Murray} {et~al.}(2021){Murray}, {Diemer}, {Chen}, {Neuhold},
  {Schnapp}, {Peruzzi}, {Blevins}, \& {Engelman}}]{murray2021}
{Murray}, S.~G., {Diemer}, B., {Chen}, Z., {et~al.} 2021, Astronomy and
  Computing, 36, 100487

\bibitem[{{Newman} {et~al.}(2024){Newman}, {Qezlou}, {Chartab}, {Rudie},
  {Blanc}, {Bird}, {Benson}, {Kelson}, \& {Lemaux}}]{newman2024}
{Newman}, A.~B., {Qezlou}, M., {Chartab}, N., {et~al.} 2024, \apj, 961, 27

\bibitem[{{Norberg} {et~al.}(2009){Norberg}, {Baugh}, {Gazta{\~n}aga}, \&
  {Croton}}]{norberg2009}
{Norberg}, P., {Baugh}, C.~M., {Gazta{\~n}aga}, E., \& {Croton}, D.~J. 2009,
  \mnras, 396, 19

\bibitem[{{Norberg} {et~al.}(2002){Norberg}, {Baugh}, {Hawkins}, {Maddox},
  {Madgwick}, {Lahav}, {Cole}, {Frenk}, {Baldry}, {Bland -Hawthorn}, {Bridges},
  {Cannon}, {Colless}, {Collins}, {Couch}, {Dalton}, {De Propris}, {Driver},
  {Efstathiou}, {Ellis}, {Glazebrook}, {Jackson}, {Lewis}, {Lumsden},
  {Peacock}, {Peterson}, {Sutherland}, \& {Taylor}}]{norberg2002}
{Norberg}, P., {Baugh}, C.~M., {Hawkins}, E., {et~al.} 2002, \mnras, 332, 827

\bibitem[{{Omori} {et~al.}(2023){Omori}, {Bottrell}, {Walmsley}, {Yesuf},
  {Goulding}, {Ding}, {Popping}, {Silverman}, {Takeuchi}, \&
  {Toba}}]{omori2023}
{Omori}, K.~C., {Bottrell}, C., {Walmsley}, M., {et~al.} 2023, \aap, 679, A142

\bibitem[{{Ostriker}(1980)}]{ostriker1980}
{Ostriker}, J.~P. 1980, Comments on Astrophysics, 8, 177

\bibitem[{{Patton} {et~al.}(2013){Patton}, {Torrey}, {Ellison}, {Mendel}, \&
  {Scudder}}]{patton2013}
{Patton}, D.~R., {Torrey}, P., {Ellison}, S.~L., {Mendel}, J.~T., \& {Scudder},
  J.~M. 2013, \mnras, 433, L59

\bibitem[{{Pearson} {et~al.}(2022){Pearson}, {Suelves}, {Ho}, {Oi}, {Brough},
  {Holwerda}, {Hopkins}, {Huang}, {Hwang}, {Kelvin}, {Kim},
  {L{\'o}pez-S{\'a}nchez}, {Ma{\l}ek}, {Pearson}, {Poliszczuk}, {Pollo},
  {Rodriguez-Gomez}, {Shim}, {Toba}, \& {Wang}}]{pearson2022}
{Pearson}, W.~J., {Suelves}, L.~E., {Ho}, S.~C.~C., {et~al.} 2022, \aap, 661,
  A52

\bibitem[{{Pearson} {et~al.}(2019{\natexlab{a}}){Pearson}, {Wang}, {Alpaslan},
  {Baldry}, {Bilicki}, {Brown}, {Grootes}, {Holwerda}, {Kitching}, {Kruk}, \&
  {van der Tak}}]{pearson2019_merger_sfr}
{Pearson}, W.~J., {Wang}, L., {Alpaslan}, M., {et~al.} 2019{\natexlab{a}},
  \aap, 631, A51

\bibitem[{{Pearson} {et~al.}(2019{\natexlab{b}}){Pearson}, {Wang}, {Trayford},
  {Petrillo}, \& {van der Tak}}]{pearson2019_identifying}
{Pearson}, W.~J., {Wang}, L., {Trayford}, J.~W., {Petrillo}, C.~E., \& {van der
  Tak}, F.~F.~S. 2019{\natexlab{b}}, \aap, 626, A49

\bibitem[{{Peebles}(1980)}]{peebles1980}
{Peebles}, P.~J.~E. 1980, {The large-scale structure of the universe}

\bibitem[{{Perez} {et~al.}(2009){Perez}, {Tissera}, {Padilla}, {Alonso}, \&
  {Lambas}}]{perez2009}
{Perez}, J., {Tissera}, P., {Padilla}, N., {Alonso}, M.~S., \& {Lambas}, D.~G.
  2009, \mnras, 399, 1157

\bibitem[{{Pollo} {et~al.}(2006){Pollo}, {Guzzo}, {Le F{\`e}vre}, {Meneux},
  {Cappi}, {Franzetti}, {Iovino}, {McCracken}, {Marinoni}, {Zamorani},
  {Bottini}, {Garilli}, {Le Brun}, {Maccagni}, {Picat}, {Scaramella},
  {Scodeggio}, {Tresse}, {Vettolani}, {Zanichelli}, {Adami}, {Arnouts},
  {Bardelli}, {Bolzonella}, {Charlot}, {Ciliegi}, {Contini}, {Foucaud},
  {Gavignaud}, {Ilbert}, {Marano}, {Mazure}, {Merighi}, {Paltani}, {Pell{\`o}},
  {Pozzetti}, {Radovich}, {Zucca}, {Bondi}, {Bongiorno}, {Busarello},
  {Cucciati}, {Gregorini}, {Lamareille}, {Mathez}, {Mellier}, {Merluzzi},
  {Ripepi}, \& {Rizzo}}]{pollo2006}
{Pollo}, A., {Guzzo}, L., {Le F{\`e}vre}, O., {et~al.} 2006, \aap, 451, 409

\bibitem[{{Press} \& {Schechter}(1974)}]{press_schechter_1974}
{Press}, W.~H. \& {Schechter}, P. 1974, \apj, 187, 425

\bibitem[{{Robaina} {et~al.}(2009){Robaina}, {Bell}, {Skelton}, {McIntosh},
  {Somerville}, {Zheng}, {Rix}, {Bacon}, {Balogh}, {Barazza}, {Barden},
  {B{\"o}hm}, {Caldwell}, {Gallazzi}, {Gray}, {H{\"a}ussler}, {Heymans},
  {Jahnke}, {Jogee}, {van Kampen}, {Lane}, {Meisenheimer}, {Papovich}, {Peng},
  {S{\'a}nchez}, {Skibba}, {Taylor}, {Wisotzki}, \& {Wolf}}]{robaina2009}
{Robaina}, A.~R., {Bell}, E.~F., {Skelton}, R.~E., {et~al.} 2009, \apj, 704,
  324

\bibitem[{{Robotham} {et~al.}(2010){Robotham}, {Driver}, {Norberg}, {Baldry},
  {Bamford}, {Hopkins}, {Liske}, {Loveday}, {Peacock}, {Cameron}, {Croom},
  {Doyle}, {Frenk}, {Hill}, {Jones}, {van Kampen}, {Kelvin}, {Kuijken},
  {Nichol}, {Parkinson}, {Popescu}, {Prescott}, {Sharp}, {Sutherland},
  {Thomas}, \& {Tuffs}}]{robotham2010_gama_tiling}
{Robotham}, A., {Driver}, S.~P., {Norberg}, P., {et~al.} 2010, \pasa, 27, 76

\bibitem[{{Roche} {et~al.}(1999){Roche}, {Eales}, {Hippelein}, \&
  {Willott}}]{roche1999}
{Roche}, N., {Eales}, S.~A., {Hippelein}, H., \& {Willott}, C.~J. 1999, \mnras,
  306, 538

\bibitem[{{Rodriguez-Gomez} {et~al.}(2019){Rodriguez-Gomez}, {Snyder}, {Lotz},
  {Nelson}, {Pillepich}, {Springel}, {Genel}, {Weinberger}, {Tacchella},
  {Pakmor}, {Torrey}, {Marinacci}, {Vogelsberger}, {Hernquist}, \&
  {Thilker}}]{rodriguez-gomez2019}
{Rodriguez-Gomez}, V., {Snyder}, G.~F., {Lotz}, J.~M., {et~al.} 2019, \mnras,
  483, 4140

\bibitem[{{Rose} {et~al.}(2023){Rose}, {Kartaltepe}, {Snyder},
  {Rodriguez-Gomez}, {Yung}, {Haro}, {Bagley}, {Calabr{\'o}}, {Cleri},
  {Cooper}, {Costantin}, {Croton}, {Dickinson}, {Finkelstein},
  {H{\"a}u{\ss}ler}, {Holwerda}, {Koekemoer}, {Kurczynski}, {Lucas}, {Mantha},
  {Papovich}, {P{\'e}rez-Gonz{\'a}lez}, {Pirzkal}, {Somerville}, {Straughn}, \&
  {Tacchella}}]{rose2023}
{Rose}, C., {Kartaltepe}, J.~S., {Snyder}, G.~F., {et~al.} 2023, \apj, 942, 54

\bibitem[{{Schweizer}(1982)}]{schweizer1982}
{Schweizer}, F. 1982, \apj, 252, 455

\bibitem[{{Shah} {et~al.}(2022){Shah}, {Kartaltepe}, {Magagnoli}, {Cox},
  {Wetherell}, {Vanderhoof}, {Cooke}, {Calabro}, {Chartab}, {Conselice},
  {Croton}, {de la Vega}, {Hathi}, {Ilbert}, {Inami}, {Kocevski}, {Koekemoer},
  {Lemaux}, {Lubin}, {Mantha}, {Marchesi}, {Martig}, {Moreno}, {Pampliega},
  {Patton}, {Salvato}, \& {Treister}}]{shah2022}
{Shah}, E.~A., {Kartaltepe}, J.~S., {Magagnoli}, C.~T., {et~al.} 2022, \apj,
  940, 4

\bibitem[{{Sheth} {et~al.}(2005){Sheth}, {Connolly}, \&
  {Skibba}}]{sheth2005_galform_models}
{Sheth}, R.~K., {Connolly}, A.~J., \& {Skibba}, R. 2005, ArXiv e-prints, astro

\bibitem[{{Sheth} {et~al.}(2006){Sheth}, {Jimenez}, {Panter}, \&
  {Heavens}}]{sheth2006}
{Sheth}, R.~K., {Jimenez}, R., {Panter}, B., \& {Heavens}, A.~F. 2006, \apjl,
  650, L25

\bibitem[{{Sheth} \& {Tormen}(2004)}]{sheth&tormen2004}
{Sheth}, R.~K. \& {Tormen}, G. 2004, \mnras, 350, 1385

\bibitem[{{Silva} {et~al.}(2018){Silva}, {Marchesini}, {Silverman}, {Skelton},
  {Iono}, {Martis}, {Marsan}, {Tadaki}, {Brammer}, \& {kartaltepe}}]{silva2018}
{Silva}, A., {Marchesini}, D., {Silverman}, J.~D., {et~al.} 2018, \apj, 868, 46

\bibitem[{{Skibba} {et~al.}(2006){Skibba}, {Sheth}, {Connolly}, \&
  {Scranton}}]{skibba2006}
{Skibba}, R., {Sheth}, R.~K., {Connolly}, A.~J., \& {Scranton}, R. 2006,
  \mnras, 369, 68

\bibitem[{{Skibba} {et~al.}(2009){Skibba}, {Bamford}, {Nichol}, {Lintott},
  {Andreescu}, {Edmondson}, {Murray}, {Raddick}, {Schawinski}, {Slosar},
  {Szalay}, {Thomas}, \& {Vandenberg}}]{skibba2009_morphology}
{Skibba}, R.~A., {Bamford}, S.~P., {Nichol}, R.~C., {et~al.} 2009, \mnras, 399,
  966

\bibitem[{{Skibba} {et~al.}(2015){Skibba}, {Coil}, {Mendez}, {Blanton}, {Bray},
  {Cool}, {Eisenstein}, {Guo}, {Miyaji}, {Moustakas}, \& {Zhu}}]{skibba2015}
{Skibba}, R.~A., {Coil}, A.~L., {Mendez}, A.~J., {et~al.} 2015, \apj, 807, 152

\bibitem[{{Skibba} {et~al.}(2013){Skibba}, {Sheth}, {Croton}, {Muldrew},
  {Abbas}, {Pearce}, \& {Shattow}}]{skibba2013}
{Skibba}, R.~A., {Sheth}, R.~K., {Croton}, D.~J., {et~al.} 2013, \mnras, 429,
  458

\bibitem[{{Skibba} {et~al.}(2014){Skibba}, {Smith}, {Coil}, {Moustakas},
  {Aird}, {Blanton}, {Bray}, {Cool}, {Eisenstein}, {Mendez}, {Wong}, \&
  {Zhu}}]{skibba2014_combining_fields}
{Skibba}, R.~A., {Smith}, M. S.~M., {Coil}, A.~L., {et~al.} 2014, \apj, 784,
  128

\bibitem[{{Snyder} {et~al.}(2015{\natexlab{a}}){Snyder}, {Lotz}, {Moody},
  {Peth}, {Freeman}, {Ceverino}, {Primack}, \& {Dekel}}]{snyder2015_structure}
{Snyder}, G.~F., {Lotz}, J., {Moody}, C., {et~al.} 2015{\natexlab{a}}, \mnras,
  451, 4290

\bibitem[{{Snyder} {et~al.}(2019){Snyder}, {Rodriguez-Gomez}, {Lotz}, {Torrey},
  {Quirk}, {Hernquist}, {Vogelsberger}, \& {Freeman}}]{snyder2019}
{Snyder}, G.~F., {Rodriguez-Gomez}, V., {Lotz}, J.~M., {et~al.} 2019, \mnras,
  486, 3702

\bibitem[{{Snyder} {et~al.}(2015{\natexlab{b}}){Snyder}, {Torrey}, {Lotz},
  {Genel}, {McBride}, {Vogelsberger}, {Pillepich}, {Nelson}, {Sales},
  {Sijacki}, {Hernquist}, \& {Springel}}]{snyder2015_morphology}
{Snyder}, G.~F., {Torrey}, P., {Lotz}, J.~M., {et~al.} 2015{\natexlab{b}},
  \mnras, 454, 1886

\bibitem[{{Solarz} {et~al.}(2015){Solarz}, {Pollo}, {Takeuchi}, {Ma{\l}ek},
  {Matsuhara}, {White}, {P{\c{e}}piak}, {Goto}, {Wada}, {Oyabu}, {Takagi},
  {Ohyama}, {Pearson}, {Hanami}, {Ishigaki}, \& {Malkan}}]{solarz2015}
{Solarz}, A., {Pollo}, A., {Takeuchi}, T.~T., {et~al.} 2015, \aap, 582, A58

\bibitem[{Stoyan \& Stoyan(1994)}]{stoyan&stoyan1994}
Stoyan, D. \& Stoyan, D.~H. 1994, Fractals, Random Shapes and Point Fields:
  Methods of Geometrical Statistics (Wiley)

\bibitem[{{Suelves} {et~al.}(2023){Suelves}, {Pearson}, \&
  {Pollo}}]{suelves2023}
{Suelves}, L.~E., {Pearson}, W.~J., \& {Pollo}, A. 2023, \aap, 669, A141

\bibitem[{{Sureshkumar} {et~al.}(2023){Sureshkumar}, {Durkalec}, {Pollo},
  {Bilicki}, {Cluver}, {Bellstedt}, {Farrow}, {Loveday}, {Taylor}, \&
  {Bland-Hawthorn}}]{sureshkumar2023}
{Sureshkumar}, U., {Durkalec}, A., {Pollo}, A., {et~al.} 2023, \aap, 669, A27

\bibitem[{{Sureshkumar} {et~al.}(2021){Sureshkumar}, {Durkalec}, {Pollo},
  {Bilicki}, {Loveday}, {Farrow}, {Holwerda}, {Hopkins}, {Liske}, {Pimbblet},
  {Taylor}, \& {Wright}}]{sureshkumar2021}
{Sureshkumar}, U., {Durkalec}, A., {Pollo}, A., {et~al.} 2021, \aap, 653, A35

\bibitem[{{Taylor} {et~al.}(2011){Taylor}, {Hopkins}, {Baldry}, {Brown},
  {Driver}, {Kelvin}, {Hill}, {Robotham}, {Bland-Hawthorn}, {Jones}, {Sharp},
  {Thomas}, {Liske}, {Loveday}, {Norberg}, {Peacock}, {Bamford}, {Brough},
  {Colless}, {Cameron}, {Conselice}, {Croom}, {Frenk}, {Gunawardhana},
  {Kuijken}, {Nichol}, {Parkinson}, {Phillipps}, {Pimbblet}, {Popescu},
  {Prescott}, {Sutherland}, {Tuffs}, {van Kampen}, \&
  {Wijesinghe}}]{taylor2011_gama_stellar_mass}
{Taylor}, E.~N., {Hopkins}, A.~M., {Baldry}, I.~K., {et~al.} 2011, \mnras, 418,
  1587

\bibitem[{{Taylor}(2005)}]{taylor2005_topcat}
{Taylor}, M.~B. 2005, in Astronomical Society of the Pacific Conference Series,
  Vol. 347, Astronomical Data Analysis Software and Systems XIV, ed.
  P.~{Shopbell}, M.~{Britton}, \& R.~{Ebert}

\bibitem[{{Tonnesen} \& {Cen}(2012)}]{tonnesen2012}
{Tonnesen}, S. \& {Cen}, R. 2012, \mnras, 425, 2313

\bibitem[{{Toomre} \& {Toomre}(1972)}]{toomre&toomre1972}
{Toomre}, A. \& {Toomre}, J. 1972, \apj, 178, 623

\bibitem[{{Tran} {et~al.}(2008){Tran}, {Moustakas}, {Gonzalez}, {Bai},
  {Zaritsky}, \& {Kautsch}}]{tran2008}
{Tran}, K.-V.~H., {Moustakas}, J., {Gonzalez}, A.~H., {et~al.} 2008, \apjl,
  683, L17

\bibitem[{{van Dokkum} {et~al.}(1999){van Dokkum}, {Franx}, {Fabricant},
  {Kelson}, \& {Illingworth}}]{vanDokkum1999}
{van Dokkum}, P.~G., {Franx}, M., {Fabricant}, D., {Kelson}, D.~D., \&
  {Illingworth}, G.~D. 1999, \apjl, 520, L95

\bibitem[{{Volonteri} {et~al.}(2003){Volonteri}, {Haardt}, \&
  {Madau}}]{volonteri2003}
{Volonteri}, M., {Haardt}, F., \& {Madau}, P. 2003, \apj, 582, 559

\bibitem[{{Walmsley} {et~al.}(2019){Walmsley}, {Ferguson}, {Mann}, \&
  {Lintott}}]{walmsley2019}
{Walmsley}, M., {Ferguson}, A. M.~N., {Mann}, R.~G., \& {Lintott}, C.~J. 2019,
  \mnras, 483, 2968

\bibitem[{{Wechsler} \& {Tinker}(2018)}]{wechsler2018}
{Wechsler}, R.~H. \& {Tinker}, J.~L. 2018, \araa, 56, 435

\bibitem[{{White} \& {Rees}(1978)}]{white&rees1978}
{White}, S.~D.~M. \& {Rees}, M.~J. 1978, \mnras, 183, 341

\bibitem[{{Wright} {et~al.}(2016){Wright}, {Robotham}, {Bourne}, {Driver},
  {Dunne}, {Maddox}, {Alpaslan}, {Andrews}, {Bauer}, {Bland-Hawthorn},
  {Brough}, {Brown}, {Clarke}, {Cluver}, {Davies}, {Grootes}, {Holwerda},
  {Hopkins}, {Jarrett}, {Kafle}, {Lange}, {Liske}, {Loveday}, {Moffett},
  {Norberg}, {Popescu}, {Smith}, {Taylor}, {Tuffs}, {Wang}, \&
  {Wilkins}}]{wright2016}
{Wright}, A.~H., {Robotham}, A.~S.~G., {Bourne}, N., {et~al.} 2016, \mnras,
  460, 765

\bibitem[{{Zehavi} {et~al.}(2011){Zehavi}, {Zheng}, {Weinberg}, {Blanton},
  {Bahcall}, {Berlind}, {Brinkmann}, {Frieman}, {Gunn}, {Lupton}, {Nichol},
  {Percival}, {Schneider}, {Skibba}, {Strauss}, {Tegmark}, \&
  {York}}]{zehavi2011}
{Zehavi}, I., {Zheng}, Z., {Weinberg}, D.~H., {et~al.} 2011, \apj, 736, 59

\bibitem[{{Zepf} \& {Koo}(1989)}]{zepf&koo1989}
{Zepf}, S.~E. \& {Koo}, D.~C. 1989, \apj, 337, 34

\bibitem[{{Zheng} {et~al.}(2007){Zheng}, {Coil}, \& {Zehavi}}]{zheng2007_HOD}
{Zheng}, Z., {Coil}, A.~L., \& {Zehavi}, I. 2007, \apj, 667, 760

\end{thebibliography}

\end{document}